\documentclass[preprint,prb,showpacs,superscriptaddress]{revtex4}

\usepackage{graphicx}
\usepackage{amsmath}

\begin{document}
\noindent
\title{Spin-Orbit Interactions in Bilayer Exciton-Condensate Ferromagnets}
 \author{Y.-P. Shim}
 \affiliation{Department of Physics, University of Texas at Austin, Austin, Texas 78712}
 \affiliation{Institute for Microstructural Sciences, National Research Council of Canada,
              Ottawa, Canada K1A 0R6}
 \author{A. H. MacDonald}
 \affiliation{Department of Physics, University of Texas at Austin, Austin, Texas 78712}
 \date{\today}

\begin{abstract}
Bilayer electron-hole systems with unequal electron and hole densities
are expected to have exciton condensate ground states with
spontaneous spin-polarization in both conduction and valence bands.
In the absence of spin-orbit and electron-hole exchange interactions
there is no coupling between the spin-orientations in the two quantum wells.
In this article we show that Rashba spin-orbit interactions lead to unconventional
magnetic anisotropies, whose strength we estimate, and
to ordered states with unusual quasiparticle spectra.
\end{abstract}


\maketitle


\section{Introduction}

An exciton is an elementary excitation of a semiconductor
in which a conduction band electron and a valence band hole form a bound state.
Like Cooper pairs in superconductors, excitons can condense under appropriate circumstances.
The broken symmetry associated with exciton condensation
is spontaneous phase coherence between conduction and valence band states.
Excitonic condensation has been one of the most anticipated forms of fermion-pair condensation
since first predicted in the early 1960's.\cite{Blatt,Keldysh,Keldysh2,Kozlov,Jerome}

Excitons are traditionally created by optically
exciting valence band electrons to the conduction band.
One key obstacle to the creation of an exciton condensate
has been the finite life time of optically generated excitons,
limited in most circumstances by optical recombination processes.
It was proposed \cite{Lozovik,Shevchenko} some time ago
that the exciton life-time can be substantially increased by spatially
separating electrons and holes in a bilayer configuration.
Progress in layered semiconductor growth techniques has now made it
feasible to realize bilayer electron-hole systems with great flexibility.
Many experimental studies have attempted to detect
condensation of optically generated excitons using photoluminescence
measurements \cite{ButovPRL,Larinov,ButovNature1,ButovNature2,SnokeNature,SnokeScience,Littlewood2004}
in coupled quantum well structures.

Systems in which electrons and holes are present in equilibrium
can be achieved in suitably gated semiconductor bilayer
systems \cite{Lilly1,Cambridge} and provide a
simpler and more ideal realization of an electron-hole fluid.
A degenerate bilayer electron-hole fluid undergoes
excitonic condensation \cite{Senatore_exciton} when the distance between the two
layers is smaller than the average distance between particles in one layer.
The interesting anomalous transport properties \cite{Eisen_AHM} associated with
exciton condensate superfluidity can be studied only in equilibrium electron-hole systems.
The essential technical development necessary to perform transport measurements
which probe exciton condensation is the perfection of techniques
which enable separate electrical contacts to electron or hole layers.\cite{Eisen_IndepContact}
Indeed magnetoexciton condensaton,
which occurs in strong magnetic fields and involves electrons and holes that can be
both in the conduction band, both in the valence band, or in separate bands,
has already been realized and studied using transport.\cite{Spielman,Kellogg,Eisen_AHM,Balatsky}
Equilibrium bilayer electron-hole fluids can be created in weak magnetic fields by
electrically generating an external potential difference between nearby
quantum wells approximately equal to the energy gap of the host semiconductor.
This paper is motivated by impressive recent progress in this direction.\cite{Lilly1,Cambridge,Lilly2,Lilly3}
One important feature of electrically generated equilibrium electron-hole fluids
is the ability to study the dependence of system properties on
the difference between electron and hole densities.
Because both conduction and valence bands are spin degenerate and condensation
is favored by perfect Fermi surface nesting,
it has been predicted that a finite population difference between electrons and holes
enriches the physics of excitonic condensates, leading in particular to
spin-polarized phases.\cite{Volkov1,Volkov2}

In this paper, we consider a double-quantum-well structure with two GaAs quantum wells
separated by an AlGaAs spacer. The larger band gap in AlGaAs compared to GaAs,
leads to potential profiles in which AlGaAs acts as a barrier
for both conduction band electrons and valence band holes.
An equilibrium electron-hole fluid is created by applying a large electric field
across the direction perpendicular to the layers so that
the valence band maximum in one quantum well can move above
than the conduction band minimum in the other.
This procedure leads to spatially separated but strongly interacting
electron and hole fluids.
In addition to making electron-hole systems stable, the external field
inevitably enhances structural inversion asymmetry
and therefore introduces (Rashba) spin-orbit(SO) interactions \cite{Rashba,Bychkov}
in both quantum wells.\cite{Nitta,Lu,Schapers,Grundler}
Recent theoretical work \cite{hakioglu1,hakioglu2} has shown that
Rashba SO interactions lead to unconventional anisotropic electron-hole pairing
in a bilayer system with equal densities of electrons and holes.
The focus of this paper is on the role the Rashba SO interaction plays in the spin-polarized phase
expected to accompany excitonic condensation whenever the electron and hole densities are unequal,
which is referred to as the exciton-condensate ferromagnet state.

%
%
\begin{figure}
    \centering
    \includegraphics[width=0.7\linewidth]{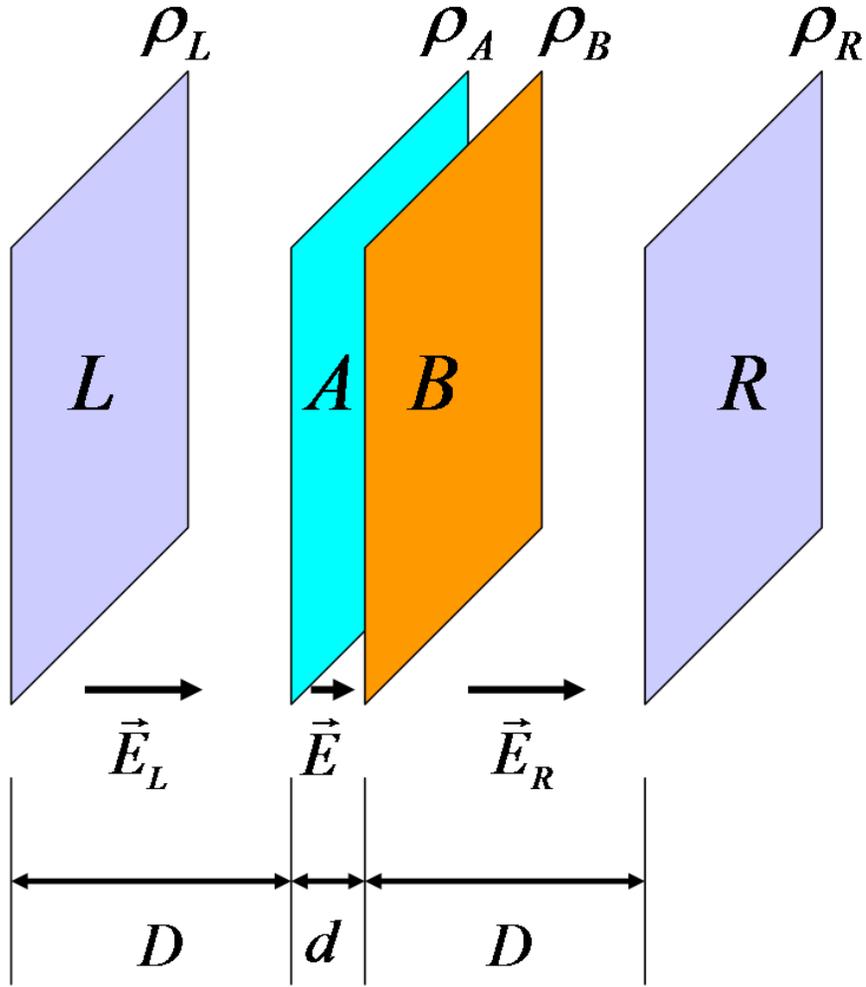}
    \caption{(Color online) Cartoon depicting the bilayer system
             including the external charge distribution
             which gives rise to the external electric field.
             Layer $A$ is the electron layer and layer $B$ is the hole layer.
             Layers $L$ and $R$ contain the external charge distribution.
             The overall charge is neutral so that
             $\rho_L + \rho_A + \rho_B + \rho_R =0$.
             $\rho_L$ and $\rho_R$ are purely external charge
             and lead to gate external fields $\mathbf{E}_L$
             and $\mathbf{E}_R$. $\rho_A=-en_e$ is contributed by the conduction
             band electrons and $\rho_B= en_h$ is the
             charge density contribution from valence band holes.
             The electric field outside of the system is zero due to the overall
             charge neutrality.}
    \label{fig:bilayer}
\end{figure}

We model the experimental system used to create the equilibrium electron-hole fluid
by assuming that external charges outside of the bilayer system
can be controlled to fix the electric fields $\mathbf{E}_L$, $\mathbf{E}_R$, on the left and right
hand sides of the bilayer system.
Treating $\mathbf{E}_L$ and $\mathbf{E}_R$, rather than the electron and hole densities,
as experimentally controllable parameters represents the real experimental condition more accurately.
The system we consider is schematically shown in Fig.~\ref{fig:bilayer}.
Two external layers ($L$ and $R$) with charge density $\rho_L$ and $\rho_R$
controls the electric field $\mathbf{E}_L$ and $\mathbf{E}_R$ which
determine the electron and hole densities in layer $A$ and $B$.
Electrons will be in layer $A$ and holes will be layer $B$.
We will use conduction-valence band picture rather than electron-hole picture
to avoid confusion in calculating the total energy of the system.
For layer $A$ ($B$), we consider only the conduction (valence) band and
neglect the valence (conduction) band.
The charge density of layer $A$ is given by $\rho_A=(-e)n_e$.
Layer $B$ also contains background positive charge with density $n_0$
so that fully occupied valence band is charge neutral.
The charge density of layer $B$ is then $\rho_B=e(n_0-n_v)\equiv e n_h$.
The overall charge neutrality is always preserved,
{\it i.e.}, $\rho_L + \rho_A + \rho_B + \rho_R =0$.
Once we fix the external charge densities $\rho_L$ and $\rho_R$,
the density difference between electrons and holes $\Delta n=n_e-n_h$
is determined by the overall charge neutrality requirement,
but $n_e$ and $n_h$ are individually determined only by requiring a constant
chemical potential through the entire system.
When $\mathbf{E}_L$ and $\mathbf{E}_R$ are identical, $\rho_L + \rho_R=0$ and therefore $n_e=n_h$,
while if $\mathbf{E}_L-\mathbf{E}_R \neq 0$, then $\rho_L + \rho_R \neq 0$
and $\Delta n = -(\rho_A +\rho_B)/e = (\rho_L+\rho_R)/e$.
Therefore, by controlling the external charge densities $\rho_L$ and $\rho_R$
or equivalently, the external electric fields $\mathbf{E}_L$ and $\mathbf{E}_R$,
it is possible to induce density-polarized ($n_e \neq n_h$) electron-hole bilayer systems.
The ability to tune the difference between electron and hole populations
is therefore a natural feature of electrically generated equilibrium electron-hole systems.
This population polarization leads to interesting new physics that is our main interest.

This paper is organized as follows.
In Sec.~\ref{sec:2band}, we briefly review the mean-field theory
of an exciton condensate for the case of spinless electrons and holes.
In Sec.~\ref{sec:4band}, we restore the spin degree of freedom
and study the spin-polarized states which occur
when the electron and hole densities are unequal.
We find that as the electron-hole population difference increases,
the ground state progresses from one with pairing in two spin-channels,
to a state with pairing in only one spin channel,
and finally to a state with no pairing.  States with finite
population polarization generally have spin-dependent pairing.
When spin-orbit interactions
and electron-hole exchange interactions
are neglected the ground state is invariant
under arbitrary spin-rotations in conduction and valence bands.
Each quantum well is spin-polarized but there is no coupling
which favors any particular relative spin orientation.
In Sec.~\ref{sec:SO} we turn to a discussion of the role played by the
Rashba spin-orbit interaction in the exciton condensate ferromagnet.
We find that spin-orbit interactions can
favor spin-alignment in a particular direction depending on the
electron and hole densities and the strength of the SO interaction.
In Sec.~\ref{sec:Summary} we briefly summarize our findings
and discuss some experimental implications of our work.

\section{Spinless Exciton Condensation}\label{sec:2band}

In this section we establish our notation and introduce the basic equations which
appear in the mean-field theory of electron-hole condensates by studying a system
with spinless conduction and valence bands localized in spatially separated quantum wells.
One of our main objectives here is to explain how we deal with the potentially
confusing way in which simple electrostatic effects are combined with pairing
in systems with electron and hole densities which are spatially separated and
controlled externally.  As it turns out, the spin degree of freedom plays a minor role
for unpolarized ($n_e=n_h$) systems, so the results in this section are applicable
with minor revisions to electron-hole bilayers with balanced populations.

We treat the charges in layer $L$ and layer $R$
as well as the background positive charge in layer $B$ as external charges.
The total Hamiltonian of a spinless bilayer system that has one conduction band
and one valence band is
\begin{equation}
\label{eq:ham}
\widehat{H} = E_{ES}^{\mathrm{ext}}
             + \sum_{a\mathbf{k}}
               \left( \varepsilon^{(0)}_{a\mathbf{k}} + V_a^\mathrm{ext} \right)
               c^{\dag}_{a\mathbf{k}} c_{a\mathbf{k}}
             + \frac{1}{2\Omega} \sum_{\substack{\mathbf{k}\mathbf{k}'\mathbf{q} \\ aa'}}
               V^{aa'}(\mathbf{q}) \,
               c^{\dag}_{a\mathbf{k}}\,c^{\dag}_{a'\mathbf{k}'}\,
               c_{a'\mathbf{k}'+\mathbf{q}}\,c_{a\mathbf{k}-\mathbf{q}}~,
\end{equation}
where the band index $a=c$ for the conduction band $c$ in layer $A$,
and $a=v$ for the valence band in layer $B$.
In Eq.~(\ref{eq:ham}) $E_{ES}^{\mathrm{ext}}$ is the electrostatic energy that comes from
the Coulomb interaction between the external charges,
and $V_a^\mathrm{ext}$ is the electrostatic potential in band $a$ due to external charges,
$\mathbf{k}$ is the two-dimensional wave vector,
$\Omega$ is area of each layer, and $V^{aa'}(\mathbf{q})$ is the Fourier transform of
the Coulomb interaction between electrons in band $a$ and band $a'$ given by
\begin{eqnarray}
V^{cc}(\mathbf{q}) &=& V^{vv}(\mathbf{q}) = \frac{2\pi e^2}{\epsilon q}~, \\
V^{cv}(\mathbf{q}) &=& V^{vc}(\mathbf{q}) = \frac{2\pi e^2}{\epsilon q} e^{-qd}~,
\end{eqnarray}
where $\epsilon$ is the dielectric constant.
In bilayer systems, unlike single-layer systems,
the Hartree electrostatic energy must be included explicitly in the mean-field theory
since individual layers are not necessarily charge neutral.
Only the overall charge neutrality is preserved.
The bare band energies $\varepsilon^{(0)}_{a\mathbf{k}}$'s are assumed to be
parabolic for both conduction and valence bands,
\begin{eqnarray}
\varepsilon^{(0)}_{c\mathbf{k}} = \frac{\hbar k^2}{2m_c}~, \qquad
\varepsilon^{(0)}_{v\mathbf{k}} = - \frac{\hbar k^2}{2m_v} -E_g ~,
\end{eqnarray}
with the fundamental band gap $E_g$.
$V_a^\mathrm{ext}$ does not depend on $\mathbf{k}$
and only depends on the $z$-coordinate of the layers.
We neglect the small overlap between electron and hole wavefunctions
so that the interaction $V$ conserves band indices and
that therefore there is no electron-hole exchange interactions.
Allowing for spontaneous inter-band coherence,
the mean-field Hartree-Fock Hamiltonian is given by
\begin{equation}\label{eq:2band_H_MF}
\widehat{H}_{MF}
 = E_{0}
  + \sum_{a\mathbf{k}}
    \left(  \varepsilon^{(0)}_{a\mathbf{k}}
          + \varepsilon^{ex}_{a\mathbf{k}}
          + V_a^\mathrm{ES} \right)
   c^{\dag}_{a\mathbf{k}} c_{a\mathbf{k}}
  -\sum_{\mathbf{k}}
   \left( \Delta_{\mathbf{k}} c^{\dag}_{c\mathbf{k}}c_{v\mathbf{k}}
         + h.c.
   \right)~,
\end{equation}
where
\begin{eqnarray}
E_{0}&=& E_{ES}-\sum_{a\mathbf{k}} V_a^\mathrm{ES}\rho^{aa}(\mathbf{k}) \nonumber\\
    &&    - \frac{1}{2} \sum_{a\mathbf{k}}
            \varepsilon^{ex}_{a\mathbf{k}}
            \rho^{aa}(\mathbf{k})
          + \frac{1}{2} \sum_{\mathbf{k}}
            \left( \Delta_{\mathbf{k}}\rho^{vc}(\mathbf{k}) + c.c. \right)~,  \\
\varepsilon^{ex}_{a\mathbf{k}}
   &=& - \frac{1}{\Omega}\sum_{\mathbf{k}'}
         V^{aa}(\mathbf{k}-\mathbf{k}')
         \rho^{aa}(\mathbf{k}')~,  \label{eq:2band_e_ex}\\
\Delta_{\mathbf{k}}
  &=& \frac{1}{\Omega}\sum_{\mathbf{k}'}
      V^{cv}(\mathbf{k}-\mathbf{k}')
      \rho^{cv}(\mathbf{k}')~, \\
\rho^{aa'}(\mathbf{k})
  &=& \langle c^{\dag}_{a'\mathbf{k}}c_{a\mathbf{k}} \rangle_{\Psi}
  = \langle \Psi | c^{\dag}_{a'\mathbf{k}}c_{a\mathbf{k}} | \Psi \rangle~.
\end{eqnarray}
Note that the Hamiltonian itself depends on the state $\Psi$ and must be solved for self-consistently.
Here $\varepsilon^{ex}_{a\mathbf{k}}$ is the intra-band exchange field,
$\Delta_{\mathbf{k}}$ is the inter-band exchange field induced by inter-band coherence,
and $\rho^{aa'}(\mathbf{k})$ is the density matrix.
These potentials, which must be also determined self-consistently, are
discussed at greater length below.
$E_{ES}$ is the electrostatic energy and
$V_a^\mathrm{ES}$ is the total electrostatic potential at layer $a$, due to
the whole charge distributions in the system
which includes the Hartree potential from the electrons and holes.
The last two terms in $E_0$ account for the double counting of
the intra- and inter-layer exchange interactions in the mean-field Hamiltonian
and must be included when we calculate the total energy.

The intra-band exchange [Eq.~\eqref{eq:2band_e_ex}] potential for the valence band
appears to diverge because $\rho^{vv}(\mathbf{k}') \to 1$ deep in the valence band.
This problem is solved by noticing that the
experimentally measured valence band effective
mass $m_v^*$ already includes all interaction effects present in the state with a
full valence band and an empty conduction band, a reference state we
will define as $|\Phi_0\rangle$.
$|\Phi_0\rangle$ is the solution of the problem with $\rho_L$=$\rho_R$=0.
Thus
\begin{eqnarray}
-\frac{\hbar k^2}{2m_v^*}
  &=& \varepsilon^{(0)}_{v\mathbf{k}} + \varepsilon^{ex}_{v\mathbf{k}}[\Phi_0]  \\
  &=& -\frac{\hbar k^2}{2m_v}
      -\frac{1}{\Omega}\sum_{\mathbf{k}'} V^{vv}(\mathbf{k}-\mathbf{k}')
       \rho_0^{vv}(\mathbf{k}')  \nonumber\\
  &=& -\frac{\hbar k^2}{2m_v}
      -\frac{1}{\Omega}\sum_{\mathbf{k}'} V^{vv}(\mathbf{k}-\mathbf{k}')
\end{eqnarray}
where $\rho_0$ is the density matrix when the system
is in state $|\Phi_0\rangle$,
\begin{equation}
\rho_0^{aa'}(\mathbf{k})
 = \left\{ \begin{array}{ll}
           1 &  a=v, a'=v \\
           0 & \textrm{otherwise}
           \end{array}
   \right.~.
\end{equation}
It follows that
\begin{equation}
\varepsilon^{(0)}_{v\mathbf{k}}+\varepsilon^{ex}_{v\mathbf{k}}
 = -\frac{\hbar k^2}{2m_v^*} - E_g
   -\frac{1}{\Omega}\sum_{\mathbf{k}'} V^{vv}(\mathbf{k}-\mathbf{k}')
    \left( \rho^{vv}(\mathbf{k}') - \rho_0^{vv}(\mathbf{k}') \right)~,
\end{equation}
which remains finite.
The total energy of the state $|\Psi \rangle$ is measured with respect to
the energy $E^{(0)}_{\mathrm{tot}}$ of the state $|\Phi_0\rangle$.

Our calculation requires a self-consistent evaluation of the difference
between the density-matrix and the density matrix in the absence of carriers:
\begin{equation}
\widetilde{\rho}^{aa'}(\mathbf{k})
  \equiv \rho^{aa'}(\mathbf{k}) -\rho_0^{aa'}(\mathbf{k})~.
\end{equation}
With this definition Eq.~\eqref{eq:2band_H_MF} becomes
\begin{eqnarray}\label{eq:2band_H_MF_new}
\widehat{H}_{MF} - E^{(0)}_{\mathrm{tot}}
&=& \widetilde{E}_{0}
  + \sum_{a\mathbf{k}}
    \varepsilon_{a\mathbf{k}}
    \left( c^{\dag}_{a\mathbf{k}} c_{a\mathbf{k}} - \rho_0^{aa}(\mathbf{k}) \right)
  - \sum_{\mathbf{k}}
    \left( \Delta_{\mathbf{k}} c^{\dag}_{c\mathbf{k}}c_{v\mathbf{k}}
         + h.c.
   \right)~,
\end{eqnarray}
where
\begin{eqnarray}
\widetilde{E}_{0}
&=& E_{ES}-E^{(0)}_{ES} -\sum_{a\mathbf{k}} V_a^\mathrm{ES}\widetilde{\rho}^{aa}(\mathbf{k}) \nonumber\\
&&    - \frac{1}{2} \sum_{a\mathbf{k}}
        \widetilde{\varepsilon}^{ex}_{a\mathbf{k}}
        \widetilde{\rho}^{aa}(\mathbf{k})
      + \frac{1}{2} \sum_{\mathbf{k}}
        \left( \Delta_{\mathbf{k}}\widetilde{\rho}^{vc}(\mathbf{k}) + c.c. \right)~, \\
\varepsilon_{a\mathbf{k}}
 &=& \widetilde{\varepsilon}^{(0)}_{a\mathbf{k}}
   + \widetilde{\varepsilon}^{ex}_{a\mathbf{k}}
   + V_a^\mathrm{ES}~, \\
\widetilde{\varepsilon}^{(0)}_{c\mathbf{k}} &=& \frac{\hbar k^2}{2m_c}~, \qquad
\widetilde{\varepsilon}^{(0)}_{v\mathbf{k}} = - \frac{\hbar k^2}{2m_v^*} -E_g ~,\\
\widetilde{\varepsilon}^{ex}_{a\mathbf{k}}
   &=& -\frac{1}{\Omega}\sum_{\mathbf{k}'} V^{aa}(\mathbf{k}-\mathbf{k}')
        \widetilde{\rho}^{aa}(\mathbf{k}')~.
\end{eqnarray}
The electrostatic energy $E^{(0)}_{\mathrm{tot}}$ of $|\Phi_0\rangle$ is zero
since there is no net charge in any of the four layers,
and the total energy $E_{\mathrm{tot}}$ of $|\Psi\rangle$ is given by
\begin{eqnarray}
E_{\mathrm{tot}} - E^{(0)}_{\mathrm{tot}}
 &=& E_{ES}
    +\sum_{a\mathbf{k}}
     \left( \widetilde{\varepsilon}_{a\mathbf{k}}^{(0)}
          + \frac{1}{2}\widetilde{\varepsilon}_{a\mathbf{k}}^{ex}
     \right)
     \widetilde{\rho}^{aa}(\mathbf{k})
    -\frac{1}{2}\sum_{a\mathbf{k}}
     \left( \Delta_{\mathbf{k}} \widetilde{\rho}^{vc}(\mathbf{k})
            + c.c. \right)~.
\end{eqnarray}
The mean-field Hamiltonian [Eq.~\eqref{eq:2band_H_MF_new}] is, in matrix form,
\begin{eqnarray}
\widehat{H}_{MF} - E^{(0)}_{\mathrm{tot}}
&=& \widetilde{E}_{0}
   -\sum_{a\mathbf{k}}
    \varepsilon_{a\mathbf{k}} \rho_0^{aa}(\mathbf{k})
   +\sum_{\mathbf{k}}
    \left( \begin{array}{cc}
           c^{\dag}_{c\mathbf{k}}&c^{\dag}_{v\mathbf{k}}
           \end{array}
    \right)
    \left( \begin{array}{cc}
           \varepsilon_{c\mathbf{k}} & -\Delta_{\mathbf{k}} \\
           -\Delta^{*}_{\mathbf{k}} & \varepsilon_{v\mathbf{k}}
           \end{array}
    \right)
    \left( \begin{array}{c}
           c_{c\mathbf{k}} \\
           c_{v\mathbf{k}}
           \end{array}
    \right)~.
\end{eqnarray}

For fixed $\rho_L$ and $\rho_R$, $\Delta n=n_e-n_h=n_L+n_R$ is fixed
and we can obtain the electrostatic potential $V_a^{ES}$ and the electrostatic energy $E_{ES}$
by solving a classical electrostatic problem of four charged layers.
The electrostatic field $V_a^{ES}$ is given by
\begin{eqnarray}
V_c^\mathrm{ES} &=& \frac{2\pi e^2 n_e d}{\epsilon} + C ~,\\
V_v^\mathrm{ES} &=&-\frac{2\pi e^2 n_h d}{\epsilon} + \varepsilon_{v0} +C~,
\end{eqnarray}
where
\begin{equation}
\varepsilon_{v0} \equiv \frac{2\pi e}{\epsilon} (\rho_L - \rho_R)d~,
\end{equation}
and $C$ is a constant reference single-particle energy which we set to zero.
The electrostatic energy per area is given up to a constant which we disregard by
\begin{equation}
\frac{E_{ES}}{\Omega} = \frac{2\pi e^2 d}{\epsilon}
\left[ \frac{1}{2}\left( n_e^2+n_h^2 \right) -n_h \left( \frac{\rho_L}{e}-\frac{\rho_R}{e} \right)\right]~.
\end{equation}

Now we derive the gap equation for the bilayer system with equal
densities of electrons and holes.
The system is completely defined by $n_L$ and $n_R$ or
alternatively by $\Delta n$ and $\mu_0\equiv \varepsilon_{v0}-E_g $.
We use the latter pair as our parameters for the following calculations.
We can diagonalize the Hamiltonian $\widehat{H}_{MF}$ by
introducing the new Fermi quasi-particle operators using a Bogoliubov
transformation:
\begin{equation}
\widehat{H}_{MF}
 = \sum_{\mathbf{k}}
   \left(\begin{array}{cc}
         \alpha^{\dag}_{\mathbf{k}} &
         \beta^{\dag}_{\mathbf{k}} \\
         \end{array}
   \right)
   \left(\begin{array}{cc}
         \varepsilon^{(1)}_{\mathbf{k}} & 0 \\
         0 & \varepsilon^{(2)}_{\mathbf{k}} \\
         \end{array}
   \right)
   \left(\begin{array}{c}
         \alpha_{\mathbf{k}} \\
         \beta_{\mathbf{k}} \\
         \end{array}
   \right)
 + \mathrm{const.}~,
\end{equation}
where
\begin{eqnarray}
\left(\begin{array}{c}
      \alpha_{\mathbf{k}} \\ \beta_{\mathbf{k}}\\
      \end{array}
\right)
  &=&
\left(\begin{array}{cc}
       u_{\mathbf{k}} & -v_{\mathbf{k}}\\
       v_{\mathbf{k}}^* & u_{\mathbf{k}}^* \\
      \end{array}
\right)
\left(\begin{array}{c}
      c_{c\mathbf{k}} \\ c_{v\mathbf{k}}
      \end{array}
\right)~, \\
\varepsilon^{(1,2)}_{\mathbf{k}}
   &=& \frac{1}{2}(\varepsilon_{c\mathbf{k}}+\varepsilon_{v\mathbf{k}})
       \pm \sqrt{\varepsilon^{2}_{\mathbf{k}}+|\Delta_{\mathbf{k}}|^{2}}~, \\
|u_{\mathbf{k}}|^2
   &=& \frac{1}{2}
       \left( 1+\frac{\varepsilon_{\mathbf{k}}}
                     {\sqrt{\varepsilon^2_{\mathbf{k}}+|\Delta_{\mathbf{k}}|^2}}
       \right)~, \\
|v_{\mathbf{k}}|^2
   &=& \frac{1}{2}
       \left( 1-\frac{\varepsilon_{\mathbf{k}}}
                     {\sqrt{\varepsilon^2_{\mathbf{k}}+|\Delta_{\mathbf{k}}|^2}}
       \right)~, \\
u_{\mathbf{k}}^* v_{\mathbf{k}}
  &=& \frac{\Delta_{\mathbf{k}}}
           {2\sqrt{\varepsilon^2_{\mathbf{k}}+|\Delta_{\mathbf{k}}|^2}}~, \\
\varepsilon_{\mathbf{k}}
  &=& \frac{1}{2} \left( \varepsilon_{c\mathbf{k}}-\varepsilon_{v\mathbf{k}} \right)~,
\end{eqnarray}
and the ground state is given by
\begin{equation}
|\Psi\rangle
   = \prod_{\mathbf{k}} \beta_{\mathbf{k}}^{\dag} |0\rangle
   = \prod_{\mathbf{k}}
     \left( v_{\mathbf{k}}c_{c\mathbf{k}}^{\dag}
          + u_{\mathbf{k}}c_{v\mathbf{k}}^{\dag}
     \right) |0\rangle~,
\end{equation}
where $|0\rangle$ is the vacuum state in which both the conduction band and the valence band are empty.
This canonical transformation is analogous to the canonical transformation of BCS theory
if we perform an electron-hole transformation on the valence band electrons.
In the conduction-valence band picture, these new fermion operators
correspond to the superposition of conduction and valence band states
that is induced by spontaneous phase coherence
and diagonalizes the self-consistent mean-field Hamiltonian.
In this language our mean-field calculation is just
standard Hartree-Fock theory.

We have described the procedure we follow in some detail
because special care is required in how the electrostatic energy is treated
in order to extract the finite energy of an electrically neutral system correctly.
We will consistently use the conduction-valence band picture of electrons
rather than the electron-hole picture throughout all the calculations for
bilayer systems because it is easier to avoid confusion in how
the cancelling Coulombic divergences are handled.
The self-consistent gap equation at zero temperature is obtained from
the definition of the inter-band exchange field,
\begin{eqnarray}\label{eq:MFT_gap}
\Delta_{\mathbf{k}}
  &=& \frac{1}{\Omega}
      \sum_{\mathbf{k}'}
      V^{cv}(\mathbf{k}-\mathbf{k}')
      \langle c^{\dag}_{v\mathbf{k}'}\,c_{c\mathbf{k}'}\rangle \nonumber\\
  &=& \frac{1}{\Omega}
      \sum_{\mathbf{k}'}
      V^{cv}(\mathbf{k}-\mathbf{k}') u_{\mathbf{k}'}^* v_{\mathbf{k}'} \nonumber\\
  &=& \frac{1}{\Omega}
      \sum_{\mathbf{k}'}
      V^{cv}(\mathbf{k}-\mathbf{k}')
      \frac{\Delta_{\mathbf{k}'}}
           {2\sqrt{\varepsilon^2_{\mathbf{k}'}+|\Delta_{\mathbf{k}'}|^2}}~.
\end{eqnarray}

%
%
\begin{figure}
    \centering
    \includegraphics[width=0.7\linewidth]{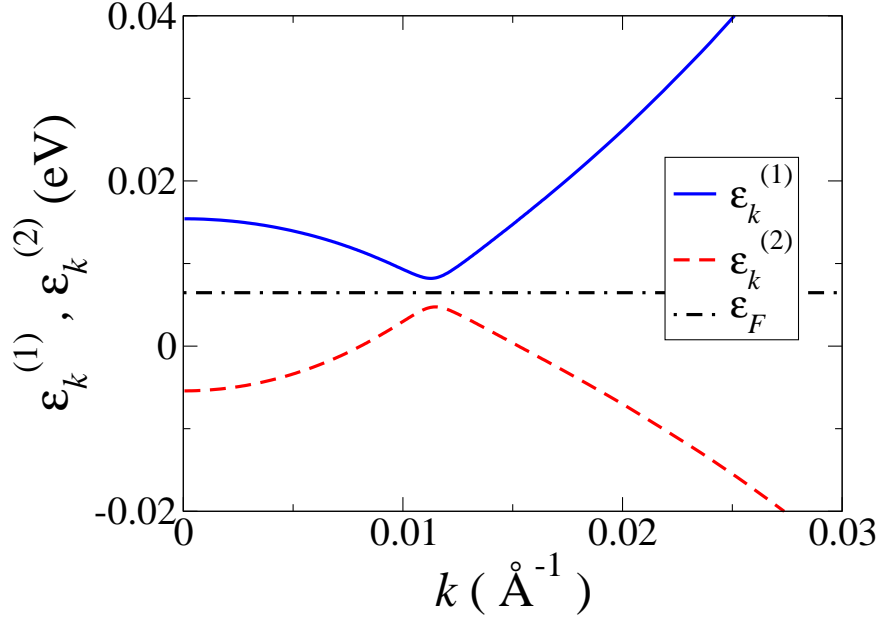} 
    \caption{(Color online) Excitation energy $\varepsilon_{\mathbf{k}}^{(1,2)}$
             of a 2-band excitonic condensate.
             $\mu_0=10 \mathrm{meV}$ and $n_e-n_h=0$.
             The calculated densities are $n_e=n_h=1.0\times 10^{11}\mathrm{cm}^{-2}$.
             $\varepsilon_F$ is the Fermi energy.}
    \label{fig:Ek_2band}
\end{figure}

%
%
\begin{figure}
    \centering
    \includegraphics[width=0.7\linewidth]{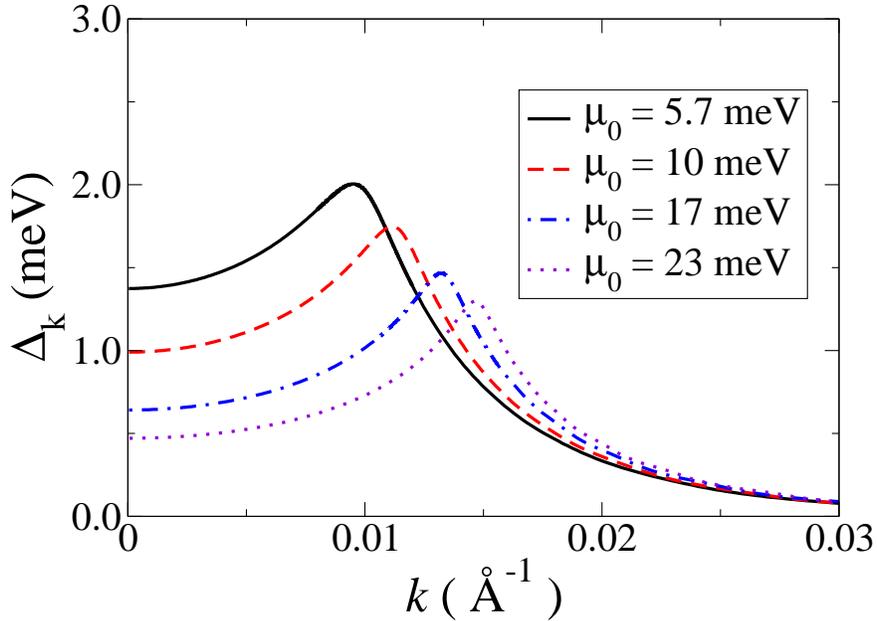}
    \caption{(Color online) $\Delta_{\mathbf{k}}$ in the 2-band model
             for various $\mu_0$ values. The calculated densities are
             $n=0.8\times 10^{11}\mathrm{cm}^{-2}$, $1.0\times 10^{11}\mathrm{cm}^{-2}$,
             $1.4\times 10^{11}\mathrm{cm}^{-2}$ and $1.7\times 10^{11}\mathrm{cm}^{-2}$
             for $\mu_0=5.7\mathrm{meV}$ to $23\mathrm{meV}$, respectively.
             }
    \label{fig:gap_total}
\end{figure}

We solved the gap equation above self-consistently for a system
with an equal density of electrons and holes,
effective masses $m_c = 0.067 m_0$, $m_v^*=0.11 m_0$,
inter-layer distance $d=100\mathrm{\AA}$, and dielectric constant $\epsilon=13$.
We choose the density difference $\Delta n = n_e - n_h$
and the initial band overlap $\mu_0 = \varepsilon_{v0}-E_g$
as two controllable parameters and set $\Delta n=0$ in this section.
There is close analogy to the BCS theory of a superconductor and we obtain
a similar energy spectrum with a gap [Fig.\ref{fig:Ek_2band}]. The gap
function $\Delta_{\mathbf{k}}$ has a maximum around the Fermi wave
vector and decreases as density increases [Fig.\ref{fig:gap_total}].
At very high densities, the ground state converges toward
the electron-hole plasma state and the gap approaches zero.
These results agree with the previous work by Zhu {\em et al}. \cite{Littlewood95}
Even though the attractive interaction is now a Coulomb interaction rather
than the BCS-like short range interaction,
the gap function shows similar behavior, which
implies that the pairing is most significant for electrons and holes
near the Fermi surface.

\section{Exciton Condensate Ferromagnets}\label{sec:4band}

We now restore the spin degree of freedom in the
bilayer systems to study its effect on excitonic condensation.
In population polarized systems this additional degree of freedom
plays an important role by allowing spontaneous spin-polarization to
compensate for imperfect nesting of the paramagnetic state Fermi surfaces.
Recently, the experimental observation of weak ferromagnetism
in lightly doped divalent hexaborides~\cite{Young} gave rise to
new interest in exciton condensate ferromagnetism as a possible explanation
for the experimental finding.\cite{Balents1,Balents2,Bascones}
Ferromagnetism in exciton condensates is driven by a preference for
Fermi surface nesting of at least one spin-component.
The spin-polarized state has some
unique symmetry properties related to the approximate spin-rotational invariance
of the microscopic Hamiltonian.
When all operators in the theory of the previous section are assigned
an additional spin index $\sigma$, the total Hamiltonian is
\begin{eqnarray}
\widehat{H}
  &=& E_{ES}^{\mathrm{ext}}
    + \sum_{a\sigma\mathbf{k}}
      \left( \varepsilon^{(0)}_{a\mathbf{k}} + V_a^{\mathrm{ext}} \right)
      c^{\dag}_{a\sigma\mathbf{k}}c_{a\sigma\mathbf{k}} \nonumber\\
 && + \frac{1}{2\Omega}
      \sum_{\substack{\mathbf{k}\mathbf{k}'\mathbf{q} \\ aa'\sigma\sigma'}}
      V(\mathbf{q})
      c^{\dag}_{a\sigma\mathbf{k}}c^{\dag}_{a'\sigma'\mathbf{k}'}
      c_{a'\sigma'\mathbf{k}'+\mathbf{q}}c_{a\sigma\mathbf{k}-\mathbf{q}}~.
\end{eqnarray}
As in the previous section, we define $|\Phi_0\rangle$ as the full-valence-band
empty-conduction-band reference state.
Electron and hole states are then specified by the density matrix
\begin{eqnarray}
\widetilde{\rho}_{\sigma\sigma'}^{aa'}(\mathbf{k})
&=& \langle \Psi |
    c^{\dag}_{a'\sigma'\mathbf{k}} c_{a\sigma\mathbf{k}}
    | \Psi \rangle
   -\rho_{0,\sigma\sigma'}^{aa'}(\mathbf{k}) \nonumber\\
&=& \langle \Psi |
    c^{\dag}_{a'\sigma'\mathbf{k}} c_{a\sigma\mathbf{k}}
    | \Psi \rangle
   -\langle \Phi_0 |
    c^{\dag}_{a'\sigma'\mathbf{k}} c_{a\sigma\mathbf{k}}
    | \Phi_0 \rangle,
\end{eqnarray}
and the mean-field Hamiltonian is
\begin{eqnarray}\label{eq:4band_H_MF}
\widehat{H}_{MF} - E^{(0)}_{\mathrm{tot}}
&=& \widetilde{E}_{0}
  + \sum_{a\sigma\mathbf{k}}
    \left(  \widetilde{\varepsilon}^{(0)}_{a\mathbf{k}}
          + \widetilde{\mathrm{h}}^{(0)}_{a\mathbf{k}}
          + V_a^\mathrm{ES} \right)
    \left( c^{\dag}_{a\sigma\mathbf{k}} c_{a\sigma\mathbf{k}}
          - \rho_{0,\sigma\sigma}^{aa}(\mathbf{k}) \right) \nonumber\\
&&+ \sum_{a\sigma\sigma'\mathbf{k}}
    \left( \widetilde{\mathbf{h}}_{a\mathbf{k}} \cdot \boldsymbol{\tau}_{\sigma\sigma'}
    \right)
    \left( c^{\dag}_{a\sigma\mathbf{k}} c_{a\sigma'\mathbf{k}}
         - \rho_{0,\sigma'\sigma}^{aa}(\mathbf{k}) \right) \nonumber\\
&&-\sum_{\mathbf{k}}
   \left( \Delta^{\sigma\sigma'}_{\mathbf{k}}
          c^{\dag}_{c\sigma\mathbf{k}} c_{v\sigma'\mathbf{k}}+ h.c.
   \right)~,
\end{eqnarray}
where the intra-band exchange field has been separated into spin-independent and spin-dependent parts:
\begin{eqnarray}
\widetilde{\mathrm{h}}_{a\mathbf{k}}^{(0)}
&=& - \frac{1}{\Omega} \sum_{\sigma\mathbf{k}'}
      V^{aa}(\mathbf{k}-\mathbf{k}')
      \widetilde{\rho}_{\sigma\sigma}^{aa}(\mathbf{k}') ~, \\
\widetilde{\mathbf{h}}_{a\mathbf{k}}
&=& - \frac{1}{2\Omega}\sum_{\sigma\sigma'\mathbf{k}'}
      V^{aa}(\mathbf{k}-\mathbf{k}')
      \widetilde{\rho}_{\sigma\sigma'}^{aa}(\mathbf{k}')
      \boldsymbol{\tau}_{\sigma'\sigma}~
\end{eqnarray}
where $\boldsymbol{\tau}$ is a vector whose components are Pauli spin
matrices.
We allow electron-hole pairing (conduction-valence coherence)
between any spin states:
\begin{eqnarray}
\Delta^{\sigma\sigma'}_{\mathbf{k}}
&=& \frac{1}{\Omega}\sum_{\mathbf{k}'}
    V^{cv}(\mathbf{k}-\mathbf{k}')
    \widetilde{\rho}_{\sigma\sigma'}^{cv}(\mathbf{k}') \label{eq:Delta}~.
\end{eqnarray}
The band energies and the electrostatic fields are then given by the same expressions
as in the spinless case
\begin{eqnarray}
\widetilde{\varepsilon}^{(0)}_{c\mathbf{k}} &=& \frac{\hbar k^2}{2m_c}~, \\
\widetilde{\varepsilon}^{(0)}_{v\mathbf{k}} &=& - \frac{\hbar k^2}{2m_v^*} -E_g ~,\\
V_c^\mathrm{ES} &=& \frac{2\pi e^2 n_e d}{\epsilon} ~,\\
V_v^\mathrm{ES} &=&-\frac{2\pi e^2 n_h d}{\epsilon} + \varepsilon_{v0}~,
\end{eqnarray}
and the constant term is
\begin{eqnarray}
\widetilde{E}_{0}
&=& E_{ES}
   -\sum_{a\sigma\mathbf{k}}
    V_a^\mathrm{ES} \widetilde{\rho}_{\sigma\sigma}^{aa}(\mathbf{k}) \nonumber\\
&& -\frac{1}{2} \sum_{a\sigma\mathbf{k}}
    \widetilde{\mathrm{h}}^{(0)}_{a\mathbf{k}}
    \widetilde{\rho}_{\sigma\sigma}^{aa}(\mathbf{k})
   -\frac{1}{2} \sum_{a\sigma\sigma'\mathbf{k}}
    \left( \widetilde{\mathbf{h}}_{a\mathbf{k}} \cdot \boldsymbol{\tau}_{\sigma\sigma'}
    \right)
    \widetilde{\rho}_{\sigma'\sigma}^{aa} (\mathbf{k})  \nonumber\\
&& +\frac{1}{2} \sum_{\sigma\sigma'\mathbf{k}}
    \left( \Delta^{\sigma\sigma'}_{\mathbf{k}}\widetilde{\rho}_{\sigma'\sigma}^{vc}(\mathbf{k})
          + c.c. \right) ~.
\end{eqnarray}
The mean-field Hamiltonian in matrix form is
\begin{equation}
\widehat{H}_{MF}
   =  \left(
      \begin{array}{cccc}
        c^{\dag}_{c\uparrow\mathbf{k}} & c^{\dag}_{c\downarrow\mathbf{k}} &
        c^{\dag}_{v\uparrow\mathbf{k}} & c^{\dag}_{v\downarrow\mathbf{k}}
      \end{array}
      \right)
      \mathbf{H}
      \left(
      \begin{array}{c}
        c_{c\uparrow\mathbf{k}} \\ c_{c\downarrow\mathbf{k}} \\
        c_{v\uparrow\mathbf{k}} \\ c_{v\downarrow\mathbf{k}} \\
      \end{array}
      \right)
    + \mathrm{const.}~,
\end{equation}
where the Hamiltonian matrix has the general form
\begin{equation}\label{eq:H_matrix}
\mathbf{H}
  = \left(
    \begin{array}{cccc}
    \varepsilon_{c\mathbf{k}} + \widetilde{\mathrm{h}}^z_{c\mathbf{k}}
      & \widetilde{\mathrm{h}}^x_{c\mathbf{k}} - i \widetilde{\mathrm{h}}^y_{c\mathbf{k}}
      & -\Delta^{\uparrow\uparrow}_{\mathbf{k}}
      & -\Delta^{\uparrow\downarrow}_{\mathbf{k}} \\
    \widetilde{\mathrm{h}}^x_{c\mathbf{k}} + i \widetilde{\mathrm{h}}^y_{c\mathbf{k}}
      & \varepsilon_{c\mathbf{k}} - \widetilde{\mathrm{h}}^z_{c\mathbf{k}}
      & -\Delta^{\downarrow\uparrow}_{\mathbf{k}}
      & -\Delta^{\downarrow\downarrow}_{\mathbf{k}} \\
    -{\Delta^*}^{\uparrow\uparrow}_{\mathbf{k}}
      & -{\Delta^*}^{\downarrow\uparrow}_{\mathbf{k}}
      & \varepsilon_{v\mathbf{k}} + \widetilde{\mathrm{h}}^z_{v\mathbf{k}}
      & \widetilde{\mathrm{h}}^x_{v\mathbf{k}} - i \widetilde{\mathrm{h}}^y_{v\mathbf{k}} \\
    -{\Delta^*}^{\uparrow\downarrow}_{\mathbf{k}}
      & -{\Delta^*}^{\downarrow\downarrow}_{\mathbf{k}}
      & \widetilde{\mathrm{h}}^x_{v\mathbf{k}} + i \widetilde{\mathrm{h}}^y_{v\mathbf{k}}
      & \varepsilon_{v\mathbf{k}} - \widetilde{\mathrm{h}}^z_{v\mathbf{k}} \\
    \end{array}
    \right)~,
\end{equation}
with
\begin{equation}
\varepsilon_{a\mathbf{k}}
  =  \widetilde{\varepsilon}_{a\mathbf{k}}^{(0)}
   + V^{\mathrm{ES}}_a
   + \widetilde{\mathrm{h}}_{a\mathbf{k}}^{(0)}~.
\end{equation}
The total energy of $\Psi$ is given by
\begin{eqnarray}
E_{\mathrm{tot}} - E^{(0)}_{\mathrm{tot}}
 &=& E_{ES}
   + \sum_{a\sigma\mathbf{k}}
     \left( \widetilde{\varepsilon}_{a\mathbf{k}}^{(0)}
          + \frac{1}{2}\widetilde{\mathrm{h}}_{a\mathbf{k}}^{(0)}
     \right)
     \widetilde{\rho}^{aa}_{\sigma\sigma}(\mathbf{k}) \nonumber\\
 && + \frac{1}{2} \sum_{a\sigma\sigma'\mathbf{k}}
      \left( \widetilde{\mathbf{h}}_{a\mathbf{k}}
             \cdot \boldsymbol{\tau}_{\sigma\sigma'}
      \right)
      \widetilde{\rho}_{\sigma'\sigma}^{aa} (\mathbf{k})
    - \frac{1}{2}\sum_{\sigma\sigma'\mathbf{k}}
      \left( \Delta^{\sigma\sigma'}_{\mathbf{k}}
             \widetilde{\rho}_{\sigma'\sigma}^{vc}(\mathbf{k})
            + c.c
      \right)~.
\end{eqnarray}

%
%
\begin{figure}
    \begin{center}
    \includegraphics[width=0.7\linewidth]{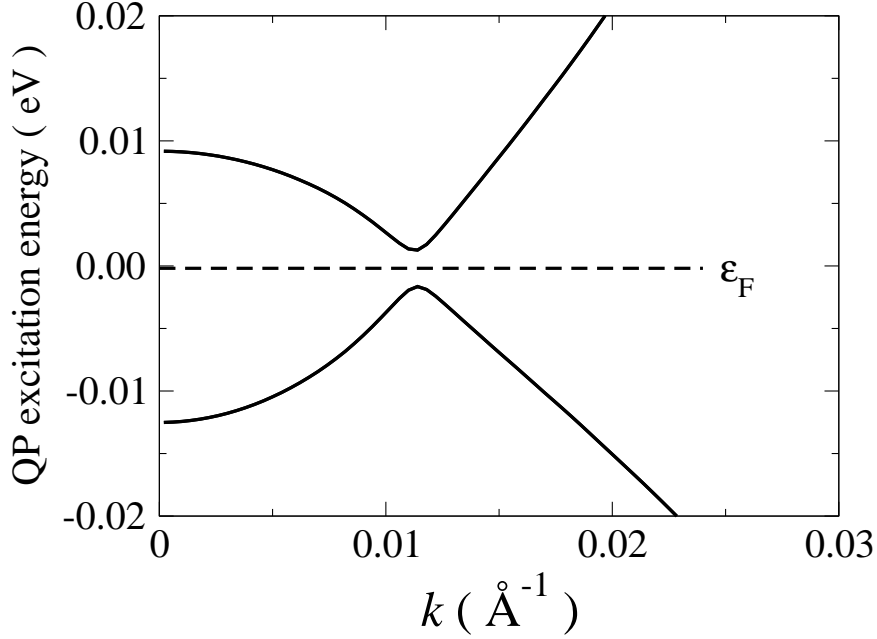}
    \caption{Quasi-particle excitation energy for
             an 4-band excitonic condensate state with
             $\mu_0=25\mathrm{meV}$ and
             $\Delta n = 0$.
             The calculated densities are
             $n_e=n_h=2.04\times10^{11}\mathrm{cm}^{-2}$. }
    \label{fig:Ek1}
    \end{center}
\end{figure}

Unlike the spinless case, the diagonalization of this $4\times 4$
matrix is not in general trivial.  In our calculations we diagonalize it
numerically and use the resulting eigenvectors to solve
the generalized Hartree-Fock equation of the system self-consistently,
allowing for both spontaneous coherence and spontaneous spin-splitting in
both conduction and valence bands.  As in the previous section, we use $\mu_0=\varepsilon_{v0}-E_g$
and $\Delta n=n_e-n_h$ as two parameters for the numerical calculations.
When $n_e = n_h$, the self-consistent calculation converges to the results of the previous section
except each quasiparticle state has an additional two-fold spin-degeneracy.
(The effective magnetic field due to the intra-band exchange term
$\widetilde{\mathbf{h}}_{a\mathbf{k}}$ vanishes in this case.)
Figure~\ref{fig:Ek1} shows the eigenvalues of the Hartree-Fock
matrix which are the quasi-particle excitation
energies. Each quasi-particle band is doubly degenerate
as expected.

\begin{figure}
    \begin{center}
    \includegraphics[width=0.9\linewidth]{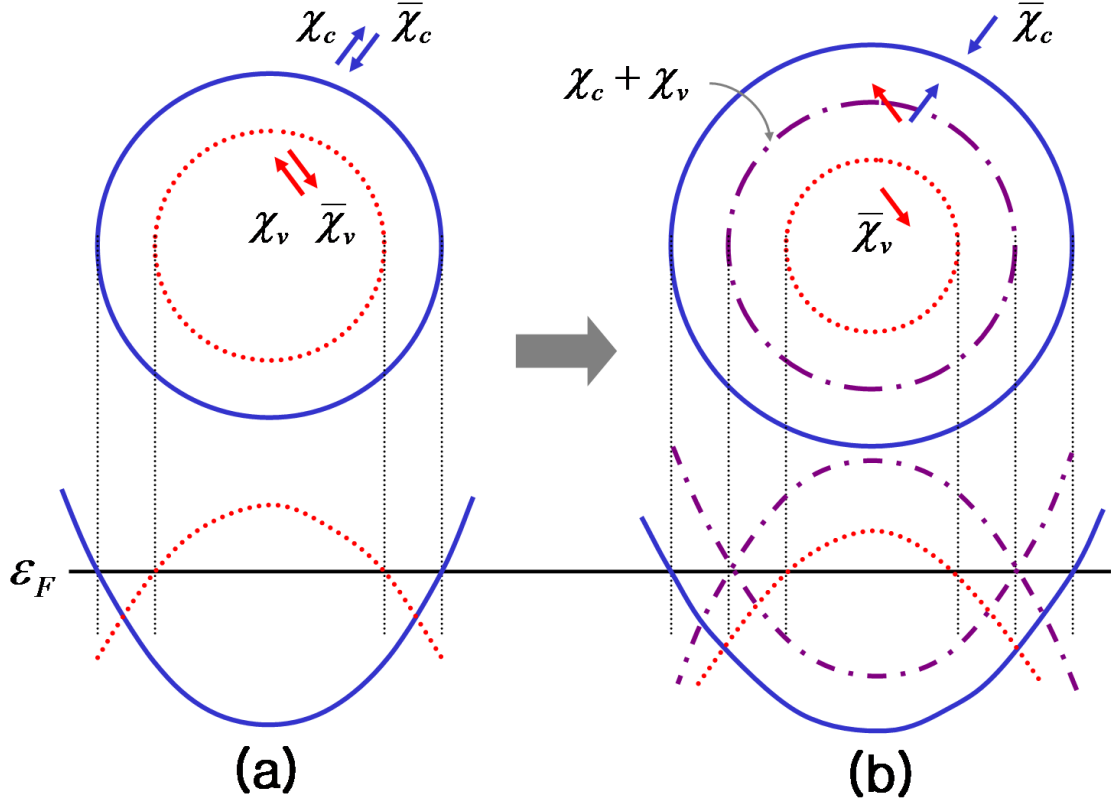}
    \caption{ (Color online) Schematic diagram of the Fermi surfaces
              and the energy bands for spontaneous
              spin splitting in ferromagnetic exciton condensates.
              (a) For different number of electrons and holes,
              the conduction band Fermi surface(solid blue circle)
              does not coincide with the valence band Fermi surface
              (dotted red circle).
              (b) Conduction band electrons with spin $\chi_c$ flip to
              spin state $\bar{\chi}_c$ so the Fermi surface of
              $\chi_c$ electrons shrinks and the Fermi surface of
              $\bar{\chi}_c$ electrons expands. Similarly, $\chi_v$ valence band electrons flip to $\bar{\chi}_v$ state
              so that the Fermi surface of the $\chi_v$ electrons increases
              until it matches the $\chi_c$ Fermi surface.
              The $\chi_c $conduction band and $\chi_v$ valence band electrons
              (dot-dashed violet circle in the middle) then condense
              to form excitonic condensates while the $\bar{\chi}_c$ and $\bar{\chi}_v$ electrons remain in the normal state.
              The spin repopulation necessary to achieve Fermi surface nesting
              leads to ferromagnetism, {\em i.e.} to spontaneous spin polarization.
              }
    \label{fig:ferromagnet}
    \end{center}
\end{figure}

When $n_e \ne n_h$
the normal state Fermi surfaces of the conduction and
valence bands have different Fermi wave vectors, frustrating
inter-band coherence which is favored by bands with
identical Fermi surfaces as we have explained.
Fermi surface matching can be restored for one conduction-valence band pair
by splitting the spin-degenerate bands so that
one conduction-valence band pair forms a condensed state
and the other pair remains in the normal state.
This spin-polarization, however,
causes the kinetic energy to increase,
which competes with the energy gained by condensation.
We therefore expect ground states with spontaneous coherence and
spontaneous spin-polarization, provided that
$|n_e - n_h|$ is not too small or too large.
We refer to these states as {\em exciton condensate ferromagnets.}
See Fig.~\ref{fig:ferromagnet} for a schematic illustration
of this point.

%
%

%
%
\begin{figure}
    \begin{center}
    \includegraphics[width=0.7\linewidth]{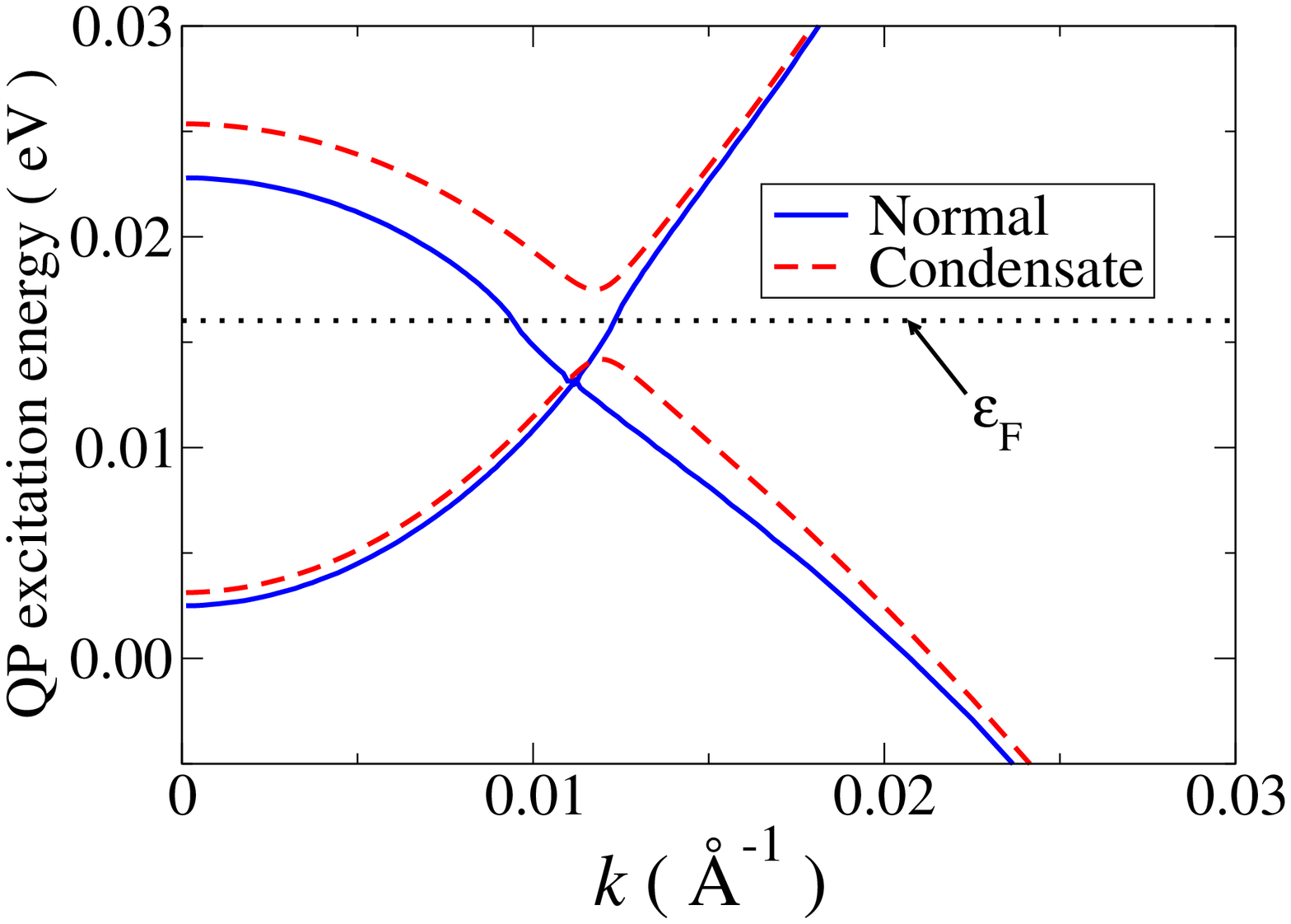}
    \caption{(Color online) Quasi-particle excitation energy for
             a ferromagnetic condensate state with
             $\mu_0=25\mathrm{meV}$ and
             $\Delta n = n_e-n_h = 5.0\times 10^{10}\mathrm{cm}^{-2}$.
             The calculated densities are
             $n_e=2.33\times10^{11}\mathrm{cm}^{-2}$
             and $n_h=1.83\times10^{11}\mathrm{cm}^{-2}$. }
    \label{fig:Ek2}
    \end{center}
\end{figure}

Figure~\ref{fig:Ek2} shows typical quasi-particle
excitation energy spectrum for a system with $n_e \ne n_h$; these results are
for $\mu_0 = 25\mathrm{meV}$ and
$\Delta n=5.0\times 10^{10}\mathrm{cm}^{-2}$.
Spontaneous splitting of the spin degeneracy indeed occurs.
One pair of bands show an energy gap due to the
condensation and the other pair show no such energy gap,
which indicates they have no interlayer coherence.
Since the valence band effective mass is larger than
the conduction band effective mass,
the splitting in the valence band is much larger than that of
the conduction band to minimize the kinetic energy cost
of achieving nesting.
The ground state energy and the energy bands are
invariant under independent spin rotations in either layer.

In bulk samples, there is a small overlap between electron and hole wave
functions which is usually neglected (dominant term approximation).
When this small overlap is not neglected, it gives rise to nonzero
electron-hole exchange interactions,
which favor spin triplet electron-hole pair states~\cite{Halperin}
and the system is invariant only under simultaneous spin-rotations.
The exciton condensate ferromagnet then behaves like a conventional
ferromagnet with negligible magnetic anisotropy.
In bilayer systems, the overlap is exponentially small
and we can therefore safely neglect electron-hole exchange for typical
inter-layer distances and barriers between the two layers.
Since we neglect the small overlap the interaction conserves the band indices
and the system has the spin-rotational symmetry for each band.
This SU(2)$\times$SU(2) symmetry leads to
degeneracy of spin singlet and triplet states.
The terms in the Hamiltonian that break this continuous symmetry
are extremely small compared to the bulk case.
Thus we have a family of ground states which differ only in the
direction of the spontaneous magnetization of each layer. This
family of ground states has the same magnitude of magnetization for
each layer but total magnetization magnitudes which vary widely.
The intra-layer interaction $\widetilde{\mathbf{h}}_{a\mathbf{k}}$
plays an important role here, helping to stabilize the ferromagnetic state
since it favors spin polarized states.  Although the spin-polarization
within each layer is driven by interlayer interactions, these do not
create a preference for the relative spin orientation of the two layers.

%
%
\begin{figure}
    \begin{center}
     \includegraphics[width=0.8\linewidth]{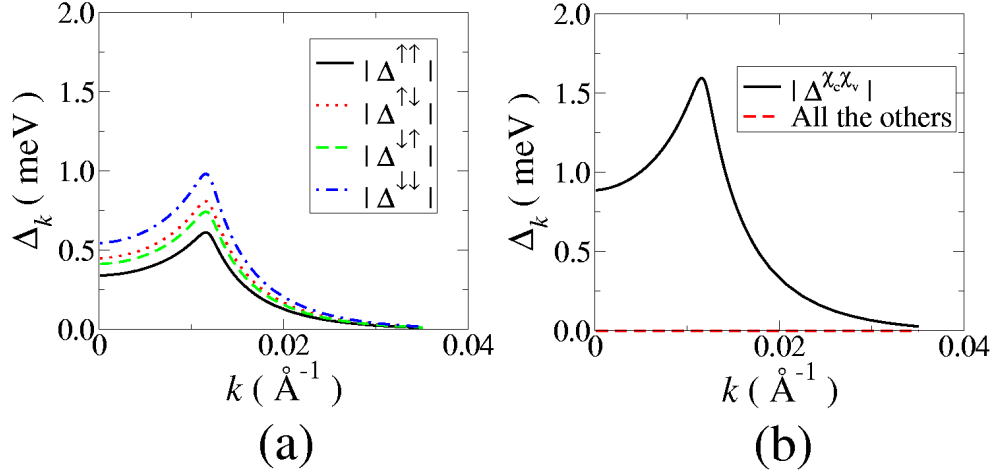}
     \caption{(Color online) The magnitudes of $\Delta_{\mathbf{k}}$ in different spin bases.
             (a) is in the spin up and down basis
             and (b) is in the new basis where
             the spin quantization direction is chosen to be
             parallel to the calculated total magnetization of each layer.
             In the new basis, the only non-vanishing $\Delta$ is
             $\Delta^{\chi_{c}\chi_{v}}$.
             This calculation is for the same model parameters used in Fig.~\ref{fig:Ek2}.
            }
    \label{fig:gap_tot}
    \end{center}
\end{figure}

In numerical calculations the
final converged spin-polarization direction depends on the initial guess we choose
to begin the self-consistent loop, and different initial states lead to different
spin-polarization directions.
To verify that the pairing occurs only between one conduction band and one valence band
leaving the other bands without coherence, we change the spin basis states from spin-up and spin-down
to $\chi_c$ and $\bar{\chi}_c$ for the conduction bands
and $\chi_v$ and $\bar{\chi}_v$ for the valence bands, which are along the spin-polarizations of
the conduction and valence bands respectively.
Let the spin-orientation for the conduction (valence) band
have polar angle $\theta_c$ ($\theta_v$) and azimuthal angle
$\phi_c$ ($\phi_v$).  Then the unitary operator $U_a^{\dag}$ that connects the two spin
basis sets is,
\begin{equation}
\left( \begin{array}{c}
       c_{a\chi_a} \\ c_{a\bar{\chi}_a}
       \end{array}
\right)
 = U_a^{\dag}
\left( \begin{array}{c}
       c_{a\uparrow} \\ c_{a\downarrow}
       \end{array}
\right)~,
\end{equation}
where
\begin{equation}
U_a^{\dag}
 = \left( \begin{array}{cc}
          \cos\frac{\theta_a}{2} & \sin\frac{\theta_a}{2}e^{-i\phi_a} \\
          \sin\frac{\theta_a}{2} & -\cos\frac{\theta_a}{2}e^{-i\phi_a}
          \end{array}
   \right)~,
\end{equation}
for $a$=$c,v$.
The order parameters in the new basis are
\begin{eqnarray}
\Delta^{\alpha_c\alpha_v}_{\mathbf{k}}
  &=& \frac{1}{\Omega}\sum_{\mathbf{k}'}
      V^{cv}(\mathbf{k}-\mathbf{k}')
      \langle c^{\dag}_{v\alpha_v\mathbf{k}'} c_{c\alpha_c\mathbf{k}'} \rangle \nonumber\\
  &=& \frac{1}{\Omega}\sum_{\mathbf{k}'}
      V^{cv}(\mathbf{k}-\mathbf{k}')
      \left(U_v^{\dag}\right)^*_{\alpha_v\sigma_v}
      \left(U_c^{\dag}\right)_{\alpha_c\sigma_c}
      \langle c^{\dag}_{v\sigma_v\mathbf{k}'} c_{c\sigma_c\mathbf{k}'} \rangle  \nonumber\\
  &=& \left(U_c^{\dag}\right)_{\alpha_c\sigma_c}
      \Delta^{\sigma_c\sigma_v}_{\mathbf{k}}
      \left(U_v\right)_{\sigma_v\alpha_v}~,
\end{eqnarray}
where $\alpha_a$ is $\chi_a$ or $\bar{\chi}_a$ and $\sigma_a$ is spin up or down.
Hence we obtain,
\begin{equation}
\left( \begin{array}{cc}
       \Delta^{\chi_c\chi_v}_{\mathbf{k}} & \Delta^{\chi_c\bar{\chi}_v}_{\mathbf{k}} \\
       \Delta^{\bar{\chi}_c\chi_v}_{\mathbf{k}} & \Delta^{\bar{\chi}_c\bar{\chi}_v}_{\mathbf{k}}
       \end{array}
\right)
 = U_c^{\dag}
\left( \begin{array}{cc}
       \Delta^{\uparrow\uparrow}_{\mathbf{k}} & \Delta^{\uparrow\downarrow}_{\mathbf{k}} \\
       \Delta^{\downarrow\uparrow}_{\mathbf{k}} & \Delta^{\downarrow\downarrow}_{\mathbf{k}}
       \end{array}
\right)
U_v~.
\end{equation}
As shown in Fig.~\ref{fig:gap_tot},
only $\Delta^{\chi_{c}\chi_{v}}$ is nonzero and all the other order
parameters are zero in the new basis,
which verifies that only $\chi_c$ and $\chi_v$ spin bands
are coherent.

The Hamiltonian can now be separated into a normal part and a condensed part:
\begin{equation}
\widehat{H}_{MF} = \widehat{H}_{N} + \widehat{H}_{C}~,
\end{equation}
where
\begin{eqnarray}
\widehat{H}_{N}
  &=& \sum_{\mathbf{k}}
      \left( \begin{array}{cc}
             c^{\dag}_{c\bar{\chi}_c\mathbf{k}} & c^{\dag}_{v\bar{\chi}_v\mathbf{k}}
             \end{array}
      \right)
      \left( \begin{array}{cc}
             \varepsilon_{c\mathbf{k}}-\widetilde{\mathrm{h}}_{c\mathbf{k}}^{\chi_c} & 0 \\
             0 & \varepsilon_{v\mathbf{k}}-\widetilde{\mathrm{h}}_{v\mathbf{k}}^{\chi_v}\\
             \end{array}
      \right)
      \left( \begin{array}{c}
             c_{c\bar{\chi}_c\mathbf{k}} \\ c_{v\bar{\chi}_v\mathbf{k}}\\
             \end{array}
      \right) ~, \label{eq:H_N} \\
\widehat{H}_{C}
  &=& \sum_{\mathbf{k}}
      \left( \begin{array}{cc}
             c^{\dag}_{c\chi_c\mathbf{k}} & c^{\dag}_{v\chi_v\mathbf{k}}
             \end{array}
      \right)
      \left( \begin{array}{cc}
             \varepsilon_{c\mathbf{k}}+\widetilde{\mathrm{h}}_{c\mathbf{k}}^{\chi_c}
             & -\Delta_{\mathbf{k}} \\
             -\Delta^*_{\mathbf{k}}
             & \varepsilon_{v\mathbf{k}}+\widetilde{\mathrm{h}}_{v\mathbf{k}}^{\chi_v}\\
             \end{array}
      \right)
      \left( \begin{array}{c}
             c_{c\chi_c\mathbf{k}} \\ c_{v\chi_v\mathbf{k}}\\
             \end{array}
       \right)~, \label{eq:H_C}
\end{eqnarray} 
and $\widetilde{\mathrm{h}}_{a\mathbf{k}}^{\chi_a}$ is the $\chi_a$-parallel component
of the intra-layer exchange field in the new spin basis, which
is the only non-vanishing component of the exchange field.
The condensate part can be diagonalized using the Bogoliubov transformation
as in the 2-band model studied in the previous section,
which leads to the eigenstates
\begin{eqnarray}
|1\mathbf{k}\rangle
  &=& |c\bar{\chi}_c \mathbf{k}\rangle~,\label{eq:new_eigenstate1}\\
|2\mathbf{k}\rangle
  &=& |v\bar{\chi}_v \mathbf{k}\rangle~,\label{eq:new_eigenstate2}\\
|3\mathbf{k}\rangle
  &=&  u^*_{\mathbf{k}}|c\chi_c \mathbf{k}\rangle
      -v^*_{\mathbf{k}} |v\chi_v \mathbf{k}\rangle~,\label{eq:new_eigenstate3}\\
|4\mathbf{k}\rangle
  &=&  v_{\mathbf{k}}|c\chi_c \mathbf{k}\rangle
      +u_{\mathbf{k}} |v\chi_v \mathbf{k}\rangle~,\label{eq:new_eigenstate4}
\end{eqnarray}
with eigenvalues
\begin{eqnarray}
\varepsilon_{1\mathbf{k}} &=& \varepsilon_{c\mathbf{k}}
                             -\widetilde{\mathrm{h}}_{c\mathbf{k}}^{\chi_c}~,\label{eq:new_eigenvalue1} \\
\varepsilon_{2\mathbf{k}} &=& \varepsilon_{v\mathbf{k}}
                             -\widetilde{\mathrm{h}}_{v\mathbf{k}}^{\chi_v}~, \label{eq:new_eigenvalue2}\\
\varepsilon_{3\mathbf{k}} &=& E^{(1)}_{\mathbf{k}}~,\label{eq:new_eigenvalue3} \\
\varepsilon_{4\mathbf{k}} &=& E^{(2)}_{\mathbf{k}}~,\label{eq:new_eigenvalue4}
\end{eqnarray}
where
\begin{eqnarray}
E^{(1,2)}_{\mathbf{k}}
 &=& \frac{1}{2}
     \left( \varepsilon_{c\mathbf{k}}+\varepsilon_{v\mathbf{k}}
           +\widetilde{\mathrm{h}}_{c\mathbf{k}}^{\chi_c}
           +\widetilde{\mathrm{h}}_{v\mathbf{k}}^{\chi_v}
     \right)
     \pm \sqrt{\varepsilon^2_{\mathbf{k}}+|\Delta_{\mathbf{k}}|^2}~, \\
\varepsilon_{\mathbf{k}}
 &=& \frac{1}{2}
     \left( \varepsilon_{c\mathbf{k}}-\varepsilon_{v\mathbf{k}}
     +\widetilde{\mathrm{h}}_{c\mathbf{k}}^{\chi_c}
     -\widetilde{\mathrm{h}}_{v\mathbf{k}}^{\chi_v}
     \right)~,
\end{eqnarray}
and
\begin{eqnarray}
|u_{\mathbf{k}}|^2
   &=& \frac{1}{2}
       \left( 1+\frac{\varepsilon_{\mathbf{k}}}
                     {\sqrt{\varepsilon^2_{\mathbf{k}}+|\Delta_{\mathbf{k}}|^2}}
       \right)~, \\
|v_{\mathbf{k}}|^2
   &=& \frac{1}{2}
       \left( 1-\frac{\varepsilon_{\mathbf{k}}}
                     {\sqrt{\varepsilon^2_{\mathbf{k}}+|\Delta_{\mathbf{k}}|^2}}
       \right) ~,\\
u_{\mathbf{k}}^* v_{\mathbf{k}}
  &=& \frac{\Delta_{\mathbf{k}}}
           {2\sqrt{\varepsilon^2_{\mathbf{k}}+|\Delta_{\mathbf{k}}|^2}} ~.
\end{eqnarray}
The ground state $|\Psi\rangle$ is given by
\begin{equation}\label{eq:Psi_G_noso}
|\Psi\rangle =
\prod_{k<k_{Fc}} c^{\dag}_{1\mathbf{k}}
\prod_{k>k_{Fv}} c^{\dag}_{2\mathbf{k}}
\prod_{\mathbf{k}} c^{\dag}_{4\mathbf{k}} \, |0\rangle ~,
\end{equation}
where $|0\rangle$ is the vacuum state with no electrons and $k_{Fc}$
($k_{Fv}$) is the Fermi wavevector for the normal conduction (valence)
band.

When the density difference is very small, the pair of bands with different Fermi surfaces
can also form a condensate and the spin repopulation readjusts to minimize the total energy.
In the other limit, if the density difference is very large, the kinetic energy cost accompanying
the spin repopulation will be larger than the condensation energy gain.
Then there will be no condensation at all and the system will remain in the normal paramagnetic state.
The states discussed above in which only one condensate occurs, are found in a broad intermediate
regime. The states which manage this compromise between different energy contributions are found
automatically by solving the mean-field equations.

\section{Influence of Rashba Spin-Orbit Interactions}\label{sec:SO}

\begin{figure}
    \begin{center}
    \includegraphics[width=0.8\linewidth]{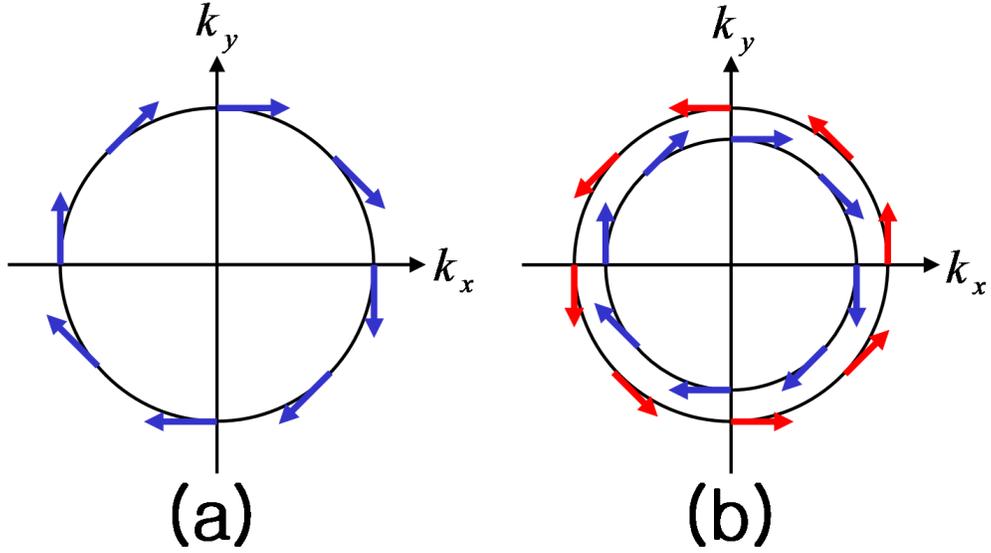}
    \caption{(Color online) (a) Rashba SO effective magnetic field
             $\mathbf{h}^{R}_{c\mathbf{k}}$
             and (b) the spin states for the conduction band.
             The spin direction is $\phi_{\mathbf{k}}-\pi/2$
             for $|c\mathbf{k}+\rangle$(inner circle and blue arrows)
             and $\phi_{\mathbf{k}}+\pi/2$
             for $|c\mathbf{k}-\rangle$(outer circle and red arrows).
             }
    \label{fig:Rashba_SO_c}
    \end{center}
\end{figure}

\begin{figure}
    \begin{center}
    \includegraphics[width=0.8\linewidth]{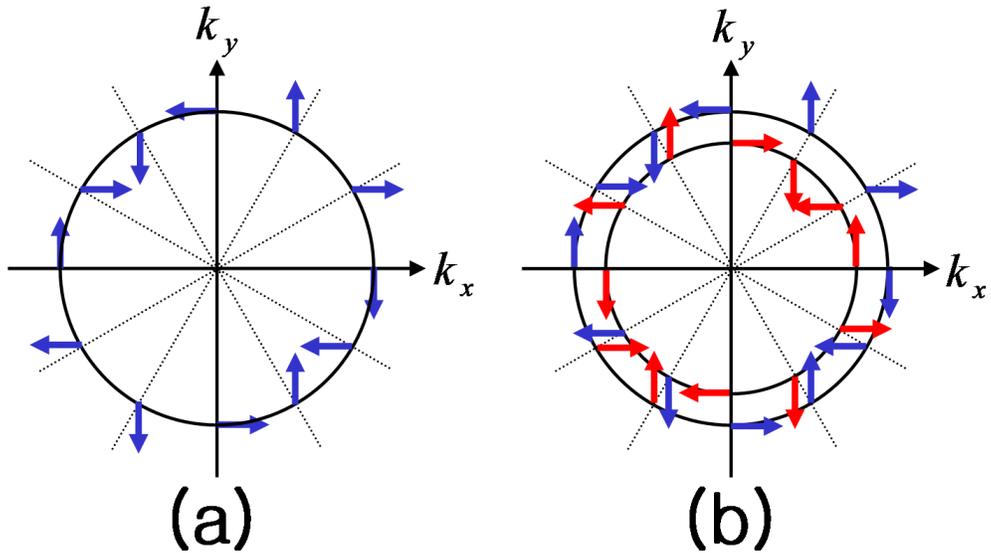}
    \caption{(Color online) (a) Rashba SO effective magnetic field
             $\mathbf{h}^{R}_{v\mathbf{k}}$
             and (b) the spin states for the heavy hole valence band.
             The spin direction is $3\phi_{\mathbf{k}}-\pi/2$
             for $|v\mathbf{k}+\rangle$(outer circle and blue arrows)
             and $3\phi_{\mathbf{k}}+\pi/2$
             for $|v\mathbf{k}-\rangle$(inner circle and red arrows).
            }
    \label{fig:Rashba_SO_hh}
    \end{center}
\end{figure}

In 2D layers, the Rashba SO interaction appears due to the structural
inversion asymmetry of the confining potential. The strength of
the SO coupling can be tuned by applying an external electric field
perpendicular to the layers.\cite{Nitta,Lu,Schapers,Grundler}
The Rashba Hamiltonian can be derived using L\"owdin perturbation
theory~\cite{Lowdin1,Lowdin2} up to the third order.\cite{Winkler_PRB02}
The effective Rashba SO interactions for the conduction band
and for the heavy hole valence band are respectively
\begin{eqnarray}
\widehat{H}^{R}_{c}
  &=& \alpha \mathbf{k} \times \hat{z} \cdot \vec{\tau} \nonumber\\
  &=& \alpha \left( \begin{array}{cc}
                    0 & ik e^{-i\phi_{\mathbf{k}}} \\
                    -ik e^{i\phi_{\mathbf{k}}} & 0 \\
                    \end{array}
             \right)\label{eq:Rashba_c} \nonumber\\
  &=& \mathbf{h}^{R}_{c\mathbf{k}} \cdot \boldsymbol{\tau}~,
\end{eqnarray}
and
\begin{eqnarray}
\widehat{H}^{R}_{hh}
  &=& i \beta (\tau_{+} k^{3}_{-} + \tau_{-} k^{3}_{+}) \nonumber\\
  &=& \beta \left( \begin{array}{cc}
                   0 & ik^{3} e^{-i 3 \phi_{\mathbf{k}}} \\
                   -ik^{3} e^{i 3 \phi_{\mathbf{k}}} & 0 \\
                   \end{array}
            \right)\label{eq:Rashba_v} \nonumber\\
  &=& \mathbf{h}^{R}_{v\mathbf{k}} \cdot \boldsymbol{\tau}~,
\end{eqnarray}
where $\tau$'s are the Pauli matrices, $\tau_{\pm}=(\tau_{x}\pm i\tau_{y}$)/2,
$k_{\pm}=k_{x}\pm i k_{y}$, and $\mathrm{tan}\phi_{\mathbf{k}}=k_{y}/k_{x}$.
These SO interactions can be considered as momentum-dependent effective magnetic fields
\begin{eqnarray}
\mathbf{h}^{R}_{c\mathbf{k}}
  &=& \alpha \left( k_y \hat{\mathbf{x}} - k_x \hat{\mathbf{y}} \right)~, \\
\mathbf{h}^{R}_{v\mathbf{k}}
  &=& \beta k^3 \left( \sin 3\phi_{\mathbf{k}}\hat{\mathbf{x}}
                     - \cos 3\phi_{\mathbf{k}}\hat{\mathbf{y}} \right)~,
\end{eqnarray}
which change direction in spin-space as directions changes in momentum space
[see Figs.~\ref{fig:Rashba_SO_c}(a) and \ref{fig:Rashba_SO_hh}(a)].
For the conduction band, the energy dispersion gets an additional linear term
$\varepsilon_{c\mathbf{k}}^{(\pm)}=\varepsilon_{c\mathbf{k}}^{(0)}\pm\alpha k$
and the corresponding spin states are
\begin{equation}
|c\mathbf{k}\pm\rangle
 = \frac{1}{\sqrt{2}}
   \left( |c\mathbf{k}\uparrow\rangle
          \mp ie^{i\phi_{\mathbf{k}}} |c\mathbf{k}\downarrow\rangle
   \right)~,
\end{equation}
which are shown in Fig.~\ref{fig:Rashba_SO_c}(b).
For the heavy hole valence band, we get an additional term proportional to $k^3$,
$\varepsilon_{v\mathbf{k}}^{(\pm)}=\varepsilon_{c\mathbf{k}}^{(0)}\pm\beta k^3$
and the corresponding spin states are
\begin{equation}
|v\mathbf{k}\pm\rangle
 = \frac{1}{\sqrt{2}}
   \left( |v\mathbf{k}\uparrow\rangle
          \mp ie^{3i\phi_{\mathbf{k}}} |v\mathbf{k}\downarrow\rangle
   \right)~,
\end{equation}
which are shown in Fig.~\ref{fig:Rashba_SO_hh}(b).

%
%
%
%

Unlike the case of the electron gas without Rashba SO interactions,
the ground state of the free electron gas does not solve the Hartree-Fock equation
of the system with Rashba SO interaction.
The spin states for each $\mathbf{k}$ are determined by the total
effective magnetic field $\mathbf{h}_{a\mathbf{k}}^{\mathrm{eff}}$
which is the sum of the intra-band exchange field
$\widetilde{\mathbf{h}}_{a\mathbf{k}}$ and the Rashba SO field $\mathbf{h}^{R}_{a\mathbf{k}}$.
For the higher energy band with energy
$\varepsilon_{a\mathbf{k}}^{(+)}=\varepsilon_{a\mathbf{k}}+|\mathbf{h}_{a\mathbf{k}}^{\mathrm{eff}}|$,
the spin is parallel to the total effective magnetic field
while for the lower energy band with energy
$\varepsilon_{a\mathbf{k}}^{(-)}=\varepsilon_{a\mathbf{k}}-|\mathbf{h}_{a\mathbf{k}}^{\mathrm{eff}}|$,
the spin state is antiparallel to the total effective magnetic field.
The self-consistency conditions for the intra-band exchange field then leads to
\begin{equation}\label{eq:self_cons_Rashba}
\widetilde{\mathbf{h}}_{a\mathbf{k}}
 = -\frac{1}{2\Omega}\sum_{\mathbf{k}'}V\left(\mathbf{k}-\mathbf{k}'\right)
    \frac{\mathbf{h}_{a\mathbf{k}'}^{\mathrm{eff}}}
         {|\mathbf{h}_{a\mathbf{k}'}^{\mathrm{eff}}|}
    \left( \widetilde{\rho}^{aa}_{++}
          -\widetilde{\rho}^{aa}_{--}
    \right)~,
\end{equation}
where the spinor of the state $|a\mathbf{k}\pm\rangle$ is parallel(+) or antiparallel(-)
to the total effective field $\mathbf{h}_{a\mathbf{k}}^{\mathrm{eff}}$,
not just the Rashba field.
The total spin $\mathbf{S}_a$ of band $a$ is given by
\begin{equation}\label{eq:Stot_Rashba}
\mathbf{S}_a
 =  \frac{1}{2}\sum_{\mathbf{k}} \;
    \frac{\mathbf{h}_{a\mathbf{k}}^{\mathrm{eff}}}
         {|\mathbf{h}_{a\mathbf{k}}^{\mathrm{eff}}|} \;
    \left( \widetilde{\rho}^{aa}_{++}(\mathbf{k}')
          -\widetilde{\rho}^{aa}_{--}(\mathbf{k}')
    \right)~.
\end{equation}
In the normal state the effect of exchange interactions is simply to enhance the
magnitude of the Rashba interaction induced spin-splitting of the bands.  For
exciton-condensate ferromagnets, however, the Rashba and exchange fields are not
in general parallel.

The Rashba SO Hamiltonian breaks spin-rotational invariance
around an arbitrary axis, but maintains invariance
under simultaneous spin and orbital rotations
around the $z$-direction,
and under inversion $z \rightarrow -z$
combined with inversion in the $xy$ plane ($\mathbf{k}\rightarrow -\mathbf{k}$).
Thus we expect that the energy of the
exciton condensate ferromagnet is independent of the azimuthal angle $\phi_a$ of the magnetization,
and invariant under reversal of the polar projection, {\em i.e.}
$\theta_a \rightarrow \pi-\theta_a$.
When the in-plane components of the Rasbha and exchange fields are not
parallel, quasiparticle energies will depend on momentum space orientation.
There will therefore tend to be some orientations in which
the conduction and valence band Fermi energies are close and
some orientations where they are more widely separated.
Condensation then occurs mainly in the region
in which the two Fermi surfaces are close together.

In the following we show that,
depending on densities and on the strength of the Rashba SO interaction,
the total spin of the ground state can be along the growth direction (uniaxial ferromagnet)
or the total spin can have nonzero in-plane component.
When the total spin is along the $z$ direction, we calculate the anisotropic energy
by applying an external magnetic field to force the total spin direction off the $z$ axis
and subtracting the magnetization energy.
When the total spin has a nonzero in-plane component,
the total spin of the ground state can have any azimuthal angle.
The rotational symmetry around the $z$ direction is spontaneously broken in this case
and the quasiparticle energy spectrum is not rotationally symmetric in $\mathbf{k}$ space.
We show representative examples for each case below.

%
%
\begin{figure}
    \begin{center}
    \includegraphics[width=0.5\linewidth]{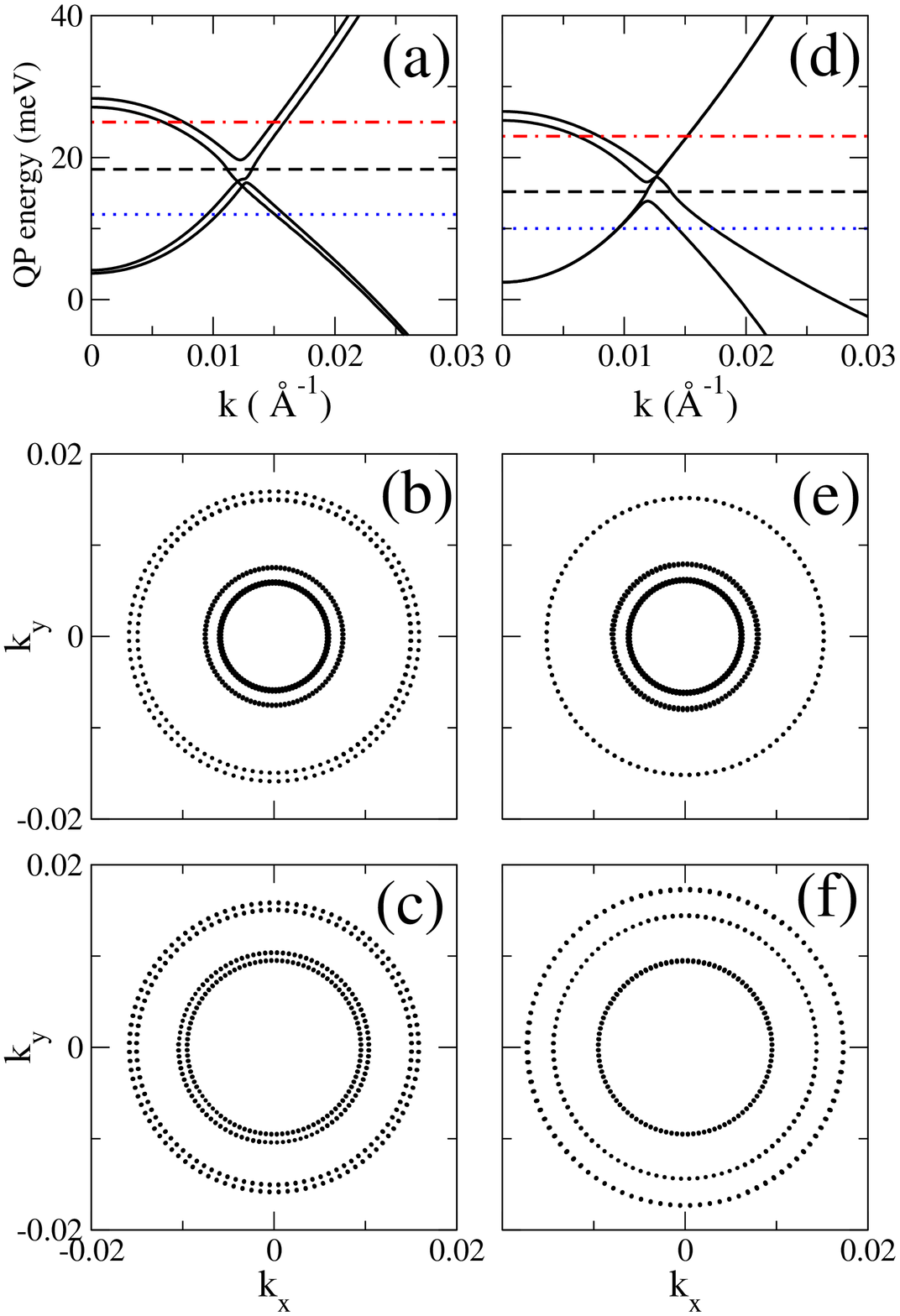}
    \caption{(Color online) Exciton condensate Ferromagnet with uniaxial magnetization
             with Rashba SO interaction in one of the layers.
             Left panels [(a)$\sim$ (c)] are for $\alpha=0.05\,\mathrm{eV \AA}$ and $\beta=0$.
             The direction of the total spin of the conduction band layer is
             along the $z$ direction.
             (a) shows the quasiparticle energy spectrum
             and the dashed black line is the chemical potential.
             (b) and (c) are constant-energy surfaces in momentum space,
             corresponding to the energies shown in (a).
             (b) is for 0.025 eV (red dot-dashed line in (a))
             and (c) is for 0.012 eV (blue dotted line in (a)).
             Right panels [(d) $\sim$ (f)] are for
             $\beta=500\,\mathrm{eV \AA^3}$ and $\alpha=0$.
             The direction of the total spin of the valence band layer
             is along the $z$ direction.
             (d) shows the quasiparticle energy spectrum
             (e) and (f) are constant-energy surfaces in momentum space.
             (e) is for 0.023 eV (red dot-dashed line in (d)) and
             (f) is for 0.01 eV (blue dotted line in (d)).
            }
    \label{fig:result1}
    \end{center}
\end{figure}

Experimental gate control should enable experimental control over the
relative strength of the Rashba interactions in valence and conduction bands.\cite{Winkler_PRB02,Winkler_PRL98}
In order to reveal the basic physics more simply
we consider systems in which the Rashba SO interaction acts only in one layer.
Figure~\ref{fig:result1} shows our results when the total spin is along the $z$ direction.
The left panels in this figure are for a system with SO interaction only in the conduction band.
The parameters are $\alpha=0.05\,\mathrm{eV \AA}$, $\beta=0$,
$\mu_0=30\mathrm{meV}$, and $\Delta n = n_e-n_h = 4.0\times 10^{10}\mathrm{cm}^{-2}$.
The densities obtained self-consistently are
$n_e=2.59\times10^{11}\mathrm{cm}^{-2}$
and $n_h=2.19\times10^{11}\mathrm{cm}^{-2}$.
The direction of the total spin of the conduction band layer is
$\theta_c=0.04\pi$ and $\phi_c=1.72\pi$, which is almost along the $z$ direction.
In this uniaxial case, the quasiparticle spectrum is
rotationally symmetric in momentum space [Fig.~\ref{fig:result1} (b) and (c)].
This circumstance is achieved in the self-consistent solution by
converging to a state in which the intra-band exchange field for the conduction band
$\widetilde{\mathbf{h}}_{c\mathbf{k}}$ has in-plane components which are
parallel to the Rashba field $\mathbf{h}^{R}_{c\mathbf{k}}$.
The total effective field of the conduction band has the form
\begin{equation}
\mathbf{h}_{c\mathbf{k}}^{\mathrm{eff}}
 = \mathrm{h}_{\bot}(k)
   \left( \sin\phi_{\mathbf{k}} \hat{\mathbf{x}}
         -\cos\phi_{\mathbf{k}} \hat{\mathbf{y}}
   \right)
   +\mathrm{h}_z(k) \hat{\mathbf{z}}~.
\end{equation}
This form implies that the magnitude of the total effective magnetic field is independent
of the momentum space orientation $\phi_{\mathbf{k}}$, explaining the isotropic Fermi surfaces.
In these solutions, the in-plane component of the total spin [Eq.~\eqref{eq:Stot_Rashba}]
for the conduction band layer vanishes and we get total magnetization $\mathbf{S}_c$ only along the $z$-direction.
On the other hand, the absence of SO interaction in the valence band leads to
arbitrary direction for the effective exchange field and therefore
the total spin $\mathbf{S}_v$ for the valence band layer is in an arbitrary direction.
Similar considerations apply for $\alpha=0$ and $\beta\neq 0$ [right panels in Fig.~\ref{fig:result1}],
where we use $\alpha=0$, $\beta=500\,\mathrm{eV \AA^3}$, $\mu_0=30\mathrm{meV}$,
and $\Delta n = n_e-n_h = -4.0\times 10^{10}\mathrm{cm}^{-2}$.
The calculated densities are
$n_e=2.24\times10^{11}\mathrm{cm}^{-2}$
and $n_h=2.64\times10^{11}\mathrm{cm}^{-2}$.
The direction of the total spin of the valence band layer
is $\theta_v=0.99\pi$ and $\phi_v=0.85\pi$.
The self-consistent solutions converge to states
with concentric constant-energy surfaces
and the total effective field is of the form
\begin{equation}
\mathbf{h}_{v\mathbf{k}}^{\mathrm{eff}}
 = \mathrm{h}_{\bot}(k)
   \left( \sin3\phi_{\mathbf{k}} \hat{\mathbf{x}}
         -\cos3\phi_{\mathbf{k}} \hat{\mathbf{y}}
   \right)
   +\mathrm{h}_z(k) \hat{\mathbf{z}}~.
\end{equation}
Again, the in-plane component of the total spin [Eq.~\eqref{eq:Stot_Rashba}] for the valence band layer
vanishes and we get total magnetization along the $z$-direction.
The direction of the total spin of the conduction band layer is arbitrary
due to the absence of SO interaction in the conduction band.

%
%
\begin{figure}
    \begin{center}
    \includegraphics[width=0.8\linewidth]{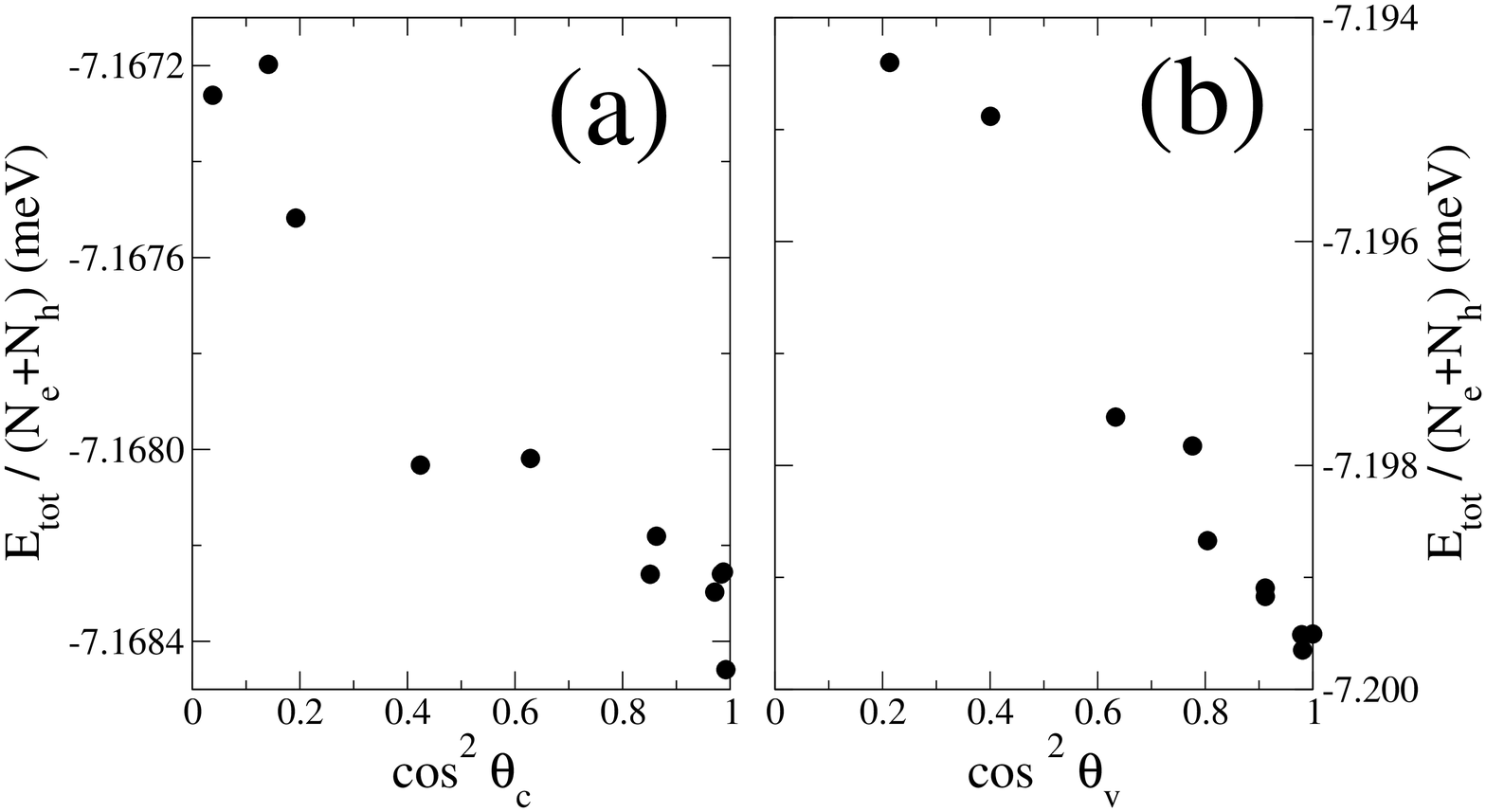}
    \caption{Magnetic anisotropy of uniaxial systems.
             $\mu_0$=$30\mathrm{meV}$ and
             $\Delta n$=$5.0\times 10^{10}\mathrm{cm}^{-2}$.
             (a)$\alpha=0.05\,\mathrm{eV \AA}$ , $\beta=0$,
             (b)$\alpha=0$, $\beta=700\,\mathrm{eV \AA^3}$.
             With these parameters, the ground state magnetization is along the $z$ direction.
             We apply small external magnetic field to change the direction of the magnetization
             and then subtract the magnetization energy to extract the energy of the system.
             The energy per particle shows a linear dependence on the $\cos^2 \theta_a$
             where $\theta_a$ is the angle between the $z$ axis and the total spin for band $a=c,v$.
             }
    \label{fig:anisotropy}
    \end{center}
\end{figure}

For these uniaxial cases, we can calculate the total energy of the system
with magnetization directions off the $z$-direction
by introducing weak external magnetic field.
Specifically, we apply an external magnetic field strong enough
so that magnetization is off the $z$ axis by $\theta_a$,
then subtract the magnetization energy $\mathbf{M}\cdot\mathbf{H}_{\mathrm{ext}}$
from the total energy to calculate the anisotropic energy of this uniaxial ferromagnet.
Figure~\ref{fig:anisotropy} shows the numerically evaluated total energy for
(a)$\alpha\neq 0$, $\beta=0$ and (b)$\alpha=0$, $\beta\neq 0$,
as a function of $\cos^2 \theta_a$.
It shows that the total energy of the uniaxial system is roughly proportional to
$\cos^2 \theta_a$.
The energy difference between the state with magnetization aligned to the $z$ axis
and the state with magnetization perpendicular to the $z$ axis
is a few $\mu$eV.
Because the magnetic anisotropy energy is very small,
Fig.~\ref{fig:anisotropy} looks somewhat noisy.
We can also derive this linear behavior by
treating the Rashba SO interaction as a perturbation (see Appendix \ref{sec:appendix} for details).
The zeroth order ground state is given by Eq.\eqref{eq:Psi_G_noso}.
We calculate the perturbed energy for each quasi-particle state
using the states Eq.~\eqref{eq:new_eigenstate1} $\sim$ Eq.~\eqref{eq:new_eigenstate4}
as the unperturbed quasi-particle states.
The total perturbed energy is evaluated by summing the corrections for each quasi-particle state
up to the Fermi energy of the unperturbed ground state, assuming that the Fermi energy
does not change much by the Rashba SO interaction.
The first order correction vanishes.  At second order we obtain
\begin{equation}
E_{\mathrm{tot}}
 = E_{\mathrm{tot}}^{(0)} + \delta E
 + A\alpha^{2}\cos^{2}\theta_{c} + B\beta^{2}\cos^{2}\theta_{v}~,
\end{equation}
where $\delta E$ is the energy correction that does not depend
on the magnetization angles and $A$, $B$ are constants.

\begin{figure}
    \begin{center}
    \includegraphics[width=0.5\linewidth]{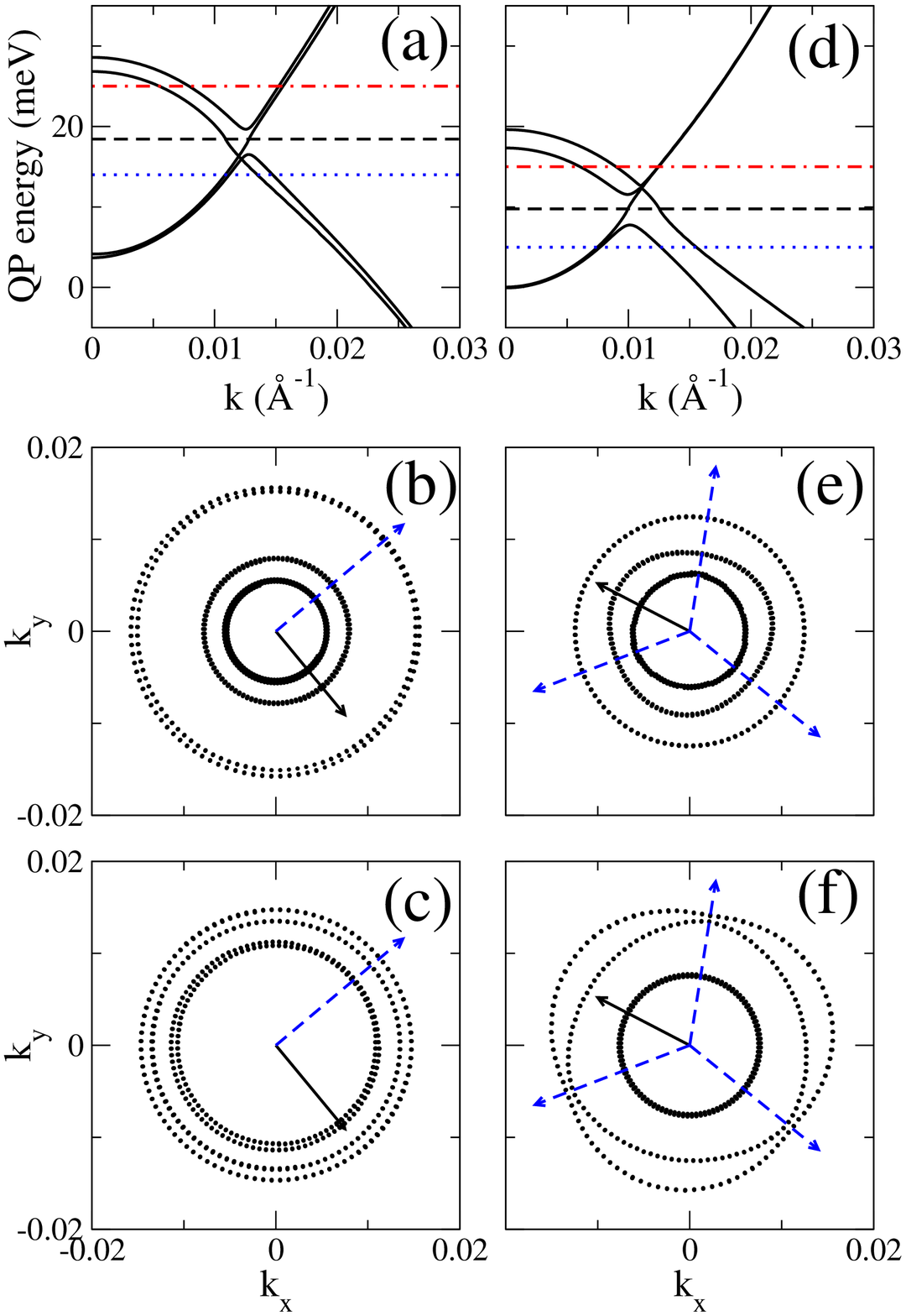}
    \caption{(Color online) Exciton condensate ferromagnet with nonzero in-plane component of
             the magnetization with Rashba SO interaction in one of the layers.
             Left panels [(a)$\sim$(c)] are for $\alpha=0.03\,\mathrm{eV \AA}$ and $\beta=0$,
             where the direction of the total spin of the conduction band layer
             is $\theta_c=0.45\pi$ and $\phi_c=1.72\pi$..
             (a) shows the quasiparticle energy spectrum
             and the dashed black line is the chemical potential.
             (b) and (c) are constant-energy surfaces in momentum space,
             corresponding to the energies shown in (a).
             (b) is for 0.025 eV (red dot-dashed line in (a))
             and (c) is for 0.014 eV (blue dotted line in (a)).
             The black solid arrow depicts the direction of the total spin
             in the conduction layer and the blue dashed arrow depicts the
             direction of $\phi_{\mathbf{k}}=\phi_c+\pi/2$.
             Right panels [(d)$\sim$(f)] are for
             $\beta=400\,\mathrm{eV \AA^3}$ and $\alpha=0$,
             where the direction of the total spin of the valence band layer
             is $\theta_v=0.63\pi$ and $\phi_v=0.85\pi$.
             (e) and (f) are constant-energy surfaces in momentum space.
             (e) is for 0.015 eV (red dot-dashed line in (d))
             and (f) is for 0.005 eV (blue dotted line in (d)).
             The black solid arrow depicts the direction of the total spin
             in the valence band layer and the blue dashed arrows
             depict the direction $\phi_{\mathbf{k}}=\phi_v/3 + \pi/6 + 2n\pi/3$
             for integer $n$.
            }
    \label{fig:result2}
    \end{center}
\end{figure}

Even though a perturbative calculation indicates
that the ferromagnet should have either an easy axis or an easy plane,
our non-perturbative self-consistent calculations sometimes find ground state with neither
an easy axis nor easy plane.
Instead, in some cases the ferromagnet can have a non-trivial optimal polar angle
and an arbitrary azimuthal angle.
Evidently higher order contributions can change the ground state qualitatively.
For a given spontaneously chosen azimuthal angle the quasi-particle band structure is anisotropic.
This broken XY symmetry leads to intricate and quite interesting quasi-particle properties.
The left panels of Fig.~\ref{fig:result2} show a case with $\alpha\neq 0$ and $\beta=0$
where the total spin of the conduction band has a nonzero in-plane component.
The parameters used for this system are $\alpha=0.03\,\mathrm{eV \AA}$, $\beta=0$,
$\mu_0=30\mathrm{meV}$, and $\Delta n = n_e-n_h = 4.0\times 10^{10}\mathrm{cm}^{-2}$.
The calculated densities are
$n_e=2.59\times10^{11}\mathrm{cm}^{-2}$
and $n_h=2.19\times10^{11}\mathrm{cm}^{-2}$.
The direction of the total spin of the conduction band layer
is $\theta_c=0.45\pi$ and $\phi_c=1.72\pi$.
As can be seen in Fig.~\ref{fig:result2}(b) and (c),
the conduction band quasi-particle excitation energy dispersions
are not rotationally symmetric.
The constant-energy surfaces for the conduction band shift so that
the two bands are closer in one direction and
farther apart in the opposite direction in $\mathbf{k}$ space.
This corresponds to an intra-band field that points in the same direction
as the Rashba field, but the magnitude of the intra-band field is
a function of not only of $k$ but also of $\phi_{\mathbf{k}}$.
To get nonzero in-plane components of the total spin along the $\phi_c$ direction,
the total effective magnetic field is stronger when the azimuthal angle of the spin states
of the majority species($|c\mathbf{k}-\rangle$ for the conduction band)
is $\phi_c$ and weaker when the azimuthal angle of the spin states of the minority
species($|c\mathbf{k}+\rangle$ for the conduction band) is $\phi_c$.
The constant-energy surfaces are farther in the direction where the effective field is stronger
and closer where the effective field is weaker since the energy difference between the two
spin bands is the magnitude of the effective magnetic field.
Therefore, we obtain the condition for the closer (farther) constant-energy surfaces
\begin{equation}
\left\{
\begin{array}{ll}
\phi_c = \phi_{\mathbf{k}} - \frac{\pi}{2};  & \textrm{closer} \\
\phi_c = \phi_{\mathbf{k}} + \frac{\pi}{2};  & \textrm{farther}
\end{array}
\right.~.
\end{equation}
The blue dashed arrows in Fig.~\ref{fig:result2}(b) and (c) show the $\phi_{\mathbf{k}}$
that satisfy the closer condition above, which agrees with the numerically calculated constant-energy
surfaces.

%
%

The right hand panels of Fig.~\ref{fig:result2} show a case with $\alpha=0$ and $\beta\neq 0$
for which the valence band magnetization has a nonzero in-plane component.
The parameters used for this calculation were $\beta=400\,\mathrm{eV \AA^3}$, $\alpha=0$,
$\mu_0=20\mathrm{meV}$, and $\Delta n = n_e-n_h = -4.0\times 10^{10}\mathrm{cm}^{-2}$.
The calculated densities are
$n_e=1.59\times10^{11}\mathrm{cm}^{-2}$
and $n_h=1.99\times10^{11}\mathrm{cm}^{-2}$.
The direction of the total spin of the valence band layer
is $\theta_v=0.63\pi$ and $\phi_v=0.85\pi$.
For the valence band, the majority species has spin state $|v\mathbf{k}-\rangle$
and minority species has $|v\mathbf{k}+\rangle$
because the valence band has opposite energy dispersion
curve compared to the conduction band.
So the majority band has spin states that point to $3\phi_{\mathbf{k}}+\pi/2$ and
the minority species has spin states that point to $3\phi_{\mathbf{k}}-\pi/2$.
Thus there are 3 closer directions and 3 farther directions that satisfy
\begin{equation}
\left\{
\begin{array}{ll}
\phi_v + 2n\pi = 3\phi_{\mathbf{k}} -\frac{\pi}{2}; & \textrm{closer} \\
\phi_v + 2n\pi = 3\phi_{\mathbf{k}} +\frac{\pi}{2}; & \textrm{farther}
\end{array}
\right.~.
\end{equation}
Fig.~\ref{fig:result2}(e) and (f) show the directions of the closer condition by blue dashed arrows,
which again agrees with the numerical results very well.
These systems spontaneously break the rotational symmetry around the $z$-direction
to make the Fermi surfaces of one conduction band and one valence band come close together
so that they can form the excitonic condensate while the other two bands remain normal.

\section{Summary and Discussion}\label{sec:Summary}

We have studied electron-hole pair condensation in spatially separated bilayer systems.
Our work is motivated by recent progress\cite{Lilly1,Lilly2,Lilly3,Cambridge} in the
preparation of electron-hole bilayers with carriers which are generated electrically
rather than optically, and in equilibrium rather than in a steady state.  These systems
already show behavior, in particular enhanced drag voltages at low-temperatures,
which appears to be evidence for a non-Fermi liquid ground state driven by
attractive interactions between conduction band electrons and valence band holes.
The extremely large
drag voltages expected\cite{Vignale} in the two-dimensional superfluid state
have not yet been seen, suggesting that the Kosterlitz-Thouless temperatures of
current samples are still below available temperatures.  More robust experimental
consequences can be anticipated if systems can be fabricated in which electron-hole interactions
have been strengthened by reducing the quantum well widths and hence the average
distance between layers.

The most unique and attractive aspects of equilibrium
electrically generated electron-hole systems are i) the opportunity to
directly probe how transport properties are altered\cite{SuNaturePhysics} by
excitonic superfluidity and ii) the opportunity to continuously
adjust the relative density of electrons and holes.
The study of pairing fermion systems with unbalanced populations\cite{coldatom}
has been a major topic in cold atom physics.  In this paper we have explored some of the
peculiarities expected to be associated with unbalanced populations in the case of
electron-hole bilayers.  The main difference between cold-atom systems and electron-hole
bilayers single-particle states have an additional attached spin-label in each layer.
This additional degree of freedom is expected to qualitatively change the
population polarization physics.
Population polarization and the subsequent Fermi surface mismatch induces spontaneous spin-polarization
which improves nesting between some Fermi surface pieces and
therefore increases the pairing condensation energy.  When spin-orbit interactions
are neglected, the spin-polarized state is invariant under independent
spin-rotations conduction valence band layers.  In this paper
we have examined how Rashba SO interactions, which will inevitably
be present in any equilibrium quantum well bilayer system, alter these
ferromagnetic condensate states.  The Rashba SO interaction
arises naturally because of the electric field used to transfer electrons between layers.
The interaction introduces a structural inversion asymmetry which breaks the spin-rotational symmetry and
leads to a preferred polar angle of the magnetization of each layer.
When an in-plane component of the magnetization is present it implies
anisotropic energy spectra and interaction driven spatial anisotropy in the two-dimensional
plane which should be readily detected experimentally.
Because the anisotropy energy is very small compared to
other energy scales, and also because of small crystalline anisotropy effects which are not
included in our calculation and appear when the spin-orbit interaction is expanded to higher order
in two-dimensional momentum, we have not been able to establish patterns in the
relationship between magnetization direction and the experimentally controllable system parameters
like the density, the density difference between electrons and holes, and the strength of the Rashba
SO interaction.

Work at the University of Texas was supported by the National Science Foundation under grant DMR-0606489.


\begin{appendix}

\section{Perturbation Theory for Exciton Condensate Ferromanget with Rashba SO Interaction}\label{sec:appendix}

%
%
The mean-field Hamiltonian of the ferromagnetic excitonic condensate of electron-hole bilayer system
without SO interaction is given by
\begin{equation}
\widehat{H}_{MF} = \widehat{H}_{N} + \widehat{H}_{C}
\end{equation}
where$\widehat{H}_{N}$ is the Hamiltonian for the normal components [Eq.~\eqref{eq:H_N}],
and $\widehat{H}_{C}$ is for the condensate [Eq.~\eqref{eq:H_C}].
For each $\mathbf{k}$, we have 4 quasiparticle eigenstates given
by Eqs.~\eqref{eq:new_eigenstate1} $\sim$ \eqref{eq:new_eigenstate4},
with eigenvalues Eqs.~\eqref{eq:new_eigenvalue1} $\sim$ \eqref{eq:new_eigenvalue4}.
These quasiparticle eigenstates are related to the conduction and valence band states by
\begin{eqnarray}
\left( \begin{array}{c}
       c_{c\uparrow\mathbf{k}}\\ c_{c\downarrow\mathbf{k}}\\
       c_{v\uparrow\mathbf{k}}\\ c_{v\downarrow\mathbf{k}}\\
       \end{array}
\right)
    &=&
\left( \begin{array}{cccc}
       \sin\frac{\theta_c}{2} & 0 &
       u^*_{\mathbf{k}}\cos\frac{\theta_c}{2} & v_{\mathbf{k}}\cos\frac{\theta_c}{2} \\
       -\cos\frac{\theta_c}{2}e^{i\phi_c} & 0 &
       u^*_{\mathbf{k}}\sin\frac{\theta_c}{2}e^{i\phi_c} & v_{\mathbf{k}}\sin\frac{\theta_c}{2}e^{i\phi_c} \\
       0 & \sin\frac{\theta_v}{2} &
       v^*_{\mathbf{k}}\cos\frac{\theta_v}{2} & u_{\mathbf{k}}\cos\frac{\theta_v}{2} \\
       0 & -\cos\frac{\theta_v}{2}e^{i\phi_v} &
       -v^*_{\mathbf{k}}\sin\frac{\theta_v}{2}e^{i\phi_v} & u_{\mathbf{k}}\sin\frac{\theta_v}{2}e^{i\phi_v} \\
       \end{array}
\right)
\left( \begin{array}{c}
       c_{1\mathbf{k}} \\ c_{2\mathbf{k}} \\ c_{3\mathbf{k}} \\ c_{4\mathbf{k}}
       \end{array}
\right)\\
    &\equiv& \mathbf{U}
\left( \begin{array}{c}
       c_{1\mathbf{k}} \\ c_{2\mathbf{k}} \\ c_{3\mathbf{k}} \\ c_{4\mathbf{k}}
       \end{array}
\right)~.
\end{eqnarray}
The ground state of this system is given by Eq.~\eqref{eq:Psi_G_noso}.
We will consider this state as our unperturbed state
and treat the Rashba SO interaction as a perturbation.

In the basis of $\{ |c\uparrow\mathbf{k}\rangle, |c\downarrow\mathbf{k}\rangle,%
|v\uparrow\mathbf{k}\rangle, |v\downarrow\mathbf{k}\rangle \}$
Rashba SO Hamiltonian is given by
\begin{equation}
\mathbf{H_R}=
\left( \begin{array}{cccc}
       0 & i\alpha ke^{-i\phi_{\mathbf{k}}} & 0 & 0 \\
       -i\alpha ke^{i\phi_{\mathbf{k}}} & 0 & 0 & 0 \\
       0 & 0 & 0 & i\beta k^3 e^{-3i\phi_{\mathbf{k}}} \\
       0 & 0 & -i\beta k^3 e^{3i\phi_{\mathbf{k}}} & 0 \\
       \end{array}
\right)~,
\end{equation}
and, in new basis $\{ |1\mathbf{k}\rangle, |2\mathbf{k}\rangle, |3\mathbf{k}\rangle, |4\mathbf{k}\rangle \}$,
by
\begin{equation}
\mathbf{H_R}^{'}=\mathbf{U}^{\dag}\mathbf{H_R}\mathbf{U}~.
\end{equation}

Applying perturbation theory for each $\mathbf{k}$, first order terms for the quasiparticle energies are
\begin{eqnarray}
\varepsilon_{1\mathbf{k}}^{(1)}&=&\left( \mathbf{H_R}^{'} \right)_{11}
                            = \alpha k\,\sin\theta_c \sin (\phi_c - \phi_{\mathbf{k}}) ~,\\
\varepsilon_{2\mathbf{k}}^{(1)}&=&\left( \mathbf{H_R}^{'} \right)_{22}
                            = \beta k^3\,\sin\theta_v \sin (\phi_v - 3\phi_{\mathbf{k}}) ~,\\
\varepsilon_{3\mathbf{k}}^{(1)}&=& \left( \mathbf{H_R}^{'} \right)_{33}
                            = -\alpha k\,|u_{\mathbf{k}}|^2\sin\theta_c \sin(\phi_c - \phi_{\mathbf{k}})
                              -\beta k^3\,|v_{\mathbf{k}}|^2\sin\theta_v \sin(\phi_v - 3\phi_{\mathbf{k}}) ~,\\
\varepsilon_{4\mathbf{k}}^{(1)}&=&\left( \mathbf{H_R}^{'} \right)_{44}
                            = -\alpha k\,|v_{\mathbf{k}}|^2\sin\theta_c \sin(\phi_c - \phi_{\mathbf{k}})
                             -\beta k^3\,|u_{\mathbf{k}}|^2\sin\theta_v \sin(\phi_v - 3\phi_{\mathbf{k}})~,
\end{eqnarray}
and the change in the total energy is
\begin{eqnarray}
\delta E_{tot}^{(1)} &=& \sum^4_{i=1}\sum_{\mathbf{k}}
                         \Theta\left(\varepsilon_F-\varepsilon^{(0)}_{i\mathbf{k}}\right)\,\varepsilon_{i\mathbf{k}}^{(1)} \nonumber\\
                     &=& \sum^4_{i=1} \frac{\Omega}{(2\pi)^2}
                         \int^{\infty}_{0} dk\,k\,\theta(\varepsilon_F-\varepsilon^{(0)}_{i\mathbf{k}})
                         \int^{2\pi}_{0} d\phi_{\mathbf{k}}\,\varepsilon_{i\mathbf{k}}^{(1)}
                      = 0~,
\end{eqnarray}
where the integration over $\phi_{\mathbf{k}}$ vanishes for all $\varepsilon_{i\mathbf{k}}^{(1)}$.
So there is no contribution from the first order terms.
Second order energy corrections for each $\mathbf{k}$ are calculated by
\begin{equation}
\varepsilon_{i\mathbf{k}}^{(2)}
  = \sum_{j \neq i} \frac{|(\mathbf{H_R}^{'})_{ij}|^2}{\varepsilon_i^{(0)}-\varepsilon_j^{(0)}}~.
\end{equation}
After some algebra, we get
\begin{eqnarray}
\varepsilon_{1\mathbf{k}}^{(2)} &=& \alpha^2 k^2
                                 \left( \frac{|u_{\mathbf{k}}|^2}{\varepsilon_{31}}
                                      + \frac{|v_{\mathbf{k}}|^2}{\varepsilon_{41}}
                                 \right)
                                 \left[
                                    \cos^2 (\phi_c - \phi_{\mathbf{k}})
                                  + \cos^2 \theta_c\,\sin^2 (\phi_c - \phi_{\mathbf{k}})
                                 \right] ~,\\
\varepsilon_{2\mathbf{k}}^{(2)} &=& \beta^2 k^6
                                 \left( \frac{|v_{\mathbf{k}}|^2}{\varepsilon_{32}}
                                      + \frac{|u_{\mathbf{k}}|^2}{\varepsilon_{42}}
                                 \right)
                                 \left[
                                    \cos^2 (\phi_v - 3\phi_{\mathbf{k}})
                                  + \cos^2 \theta_v\,\sin^2 (\phi_v - 3\phi_{\mathbf{k}})
                                 \right]~,\\
\varepsilon_{3\mathbf{k}}^{(2)} &=& \frac{\alpha^2 k^2 |u_{\mathbf{k}}|^2}{\varepsilon_{13}}
                                 \left[
                                     \cos^2 (\phi_c - \phi_{\mathbf{k}})
                                   + \cos^2 \theta_c\,\sin^2 (\phi_c - \phi_{\mathbf{k}})
                                 \right] \nonumber\\
                               &+& \frac{\beta^2 k^6 |v_{\mathbf{k}}|^2}{\varepsilon_{23}}
                                 \left[
                                     \cos^2 (\phi_v - 3\phi_{\mathbf{k}})
                                   + \cos^2 \theta_v\,\sin^2 (\phi_v - 3\phi_{\mathbf{k}})
                                 \right]\nonumber\\
                              &+&\frac{|u_{\mathbf{k}} v_{\mathbf{k}}|^2}{\varepsilon_{43}}
                                 \Big[  \alpha^2 k^2 \sin^2 \theta_c \sin^2(\phi_c - \phi_{\mathbf{k}})
                                   +\beta^2 k^6  \sin^2 \theta_v \sin^2(\phi_v - 3\phi_{\mathbf{k}})  \nonumber\\
                   &&\qquad\qquad  -2\alpha\beta\,k^4 \sin\theta_c \sin\theta_v
                                    \sin(\phi_c - \phi_{\mathbf{k}}) \sin(\phi_v - 3\phi_{\mathbf{k}})
                                  \Big]~,\\
\varepsilon_{4\mathbf{k}}^{(2)} &=&  \frac{\alpha^2 k^2 |v_{\mathbf{k}}|^2}{\varepsilon_{14}}
                                 \left[
                                     \cos^2 (\phi_c - \phi_{\mathbf{k}})
                                   + \cos^2 \theta_c\,\sin^2 (\phi_c - \phi_{\mathbf{k}})
                                 \right] \nonumber\\
                               &+& \frac{\beta^2 k^6 |v_{\mathbf{k}}|^2}{\varepsilon_{24}}
                                 \left[
                                     \cos^2 (\phi_v - 3\phi_{\mathbf{k}})
                                   + \cos^2 \theta_v\,\sin^2 (\phi_v - 3\phi_{\mathbf{k}})
                                 \right]\nonumber\\
                              &+&\frac{|u_{\mathbf{k}} v_{\mathbf{k}}|^2}{\varepsilon_{34}}
                                 \Big[
                                    \alpha^2 k^2 \sin^2 \theta_c \sin^2(\phi_c - \phi_{\mathbf{k}})
                                   +\beta^2 k^6  \sin^2 \theta_v \sin^2(\phi_v - 3\phi_{\mathbf{k}})
                                   \nonumber\\
                  && \qquad\qquad -2\alpha\beta\,k^4 \sin\theta_c \sin\theta_v
                                    \sin(\phi_c - \phi_{\mathbf{k}}) \sin(\phi_v - 3\phi_{\mathbf{k}})
                                  \Big]~,
\end{eqnarray}
where $\varepsilon_{ij}=\varepsilon^{(0)}_{j}-\varepsilon^{(0)}_{i}$.
Using
\begin{eqnarray}
&& \int^{2\pi}_{0} \, d\phi_{\mathbf{k}} \, \sin^2(\phi_c-\phi_{\mathbf{k}})
=\int^{2\pi}_{0} \, d\phi_{\mathbf{k}} \, \cos^2(\phi_c-\phi_{\mathbf{k}})
= \pi ~,\\
&& \int^{2\pi}_{0} \, d\phi_{\mathbf{k}} \, \sin^2(\phi_v-3\phi_{\mathbf{k}})
=\int^{2\pi}_{0} \, d\phi_{\mathbf{k}} \, \cos^2(\phi_v-3\phi_{\mathbf{k}})
= \pi ~,\\
&& \int^{2\pi}_{0} \, d\phi_{\mathbf{k}} \, \sin(\phi_c-\phi_{\mathbf{k}}) \sin(\phi_v-3\phi_{\mathbf{k}})
= 0~,
\end{eqnarray}
we can find second order energy correction
\begin{eqnarray}
\delta E_{tot}^{(2)} &=& \sum_{i=1}^4 \sum_{\mathbf{k}}
                         \Theta\left(\varepsilon_F-\varepsilon^{(0)}_{i\mathbf{k}}\right)\,\varepsilon_{i\mathbf{k}}^{(2)} \nonumber\\
                     &=& \sum^4_{i=1} \frac{\Omega}{(2\pi)^2}
                         \int^{\infty}_{0} dk\,k\,\theta(\varepsilon_F-\varepsilon^{(0)}_{i\mathbf{k}})
                         \int^{2\pi}_{0} d\phi_{\mathbf{k}}\,\varepsilon_{i\mathbf{k}}^{(2)} \nonumber\\
                     &=& \frac{\Omega}{4\pi}
                         \int^{k_{fc}}_{0} dk \alpha^2 k^3
                         \left( \frac{|u_{\mathbf{k}}|^2}{\varepsilon_{31}}
                              + \frac{|v_{\mathbf{k}}|^2}{\varepsilon_{41}}
                         \right)
                         \left( 1+ \cos^2 \theta_c \right) \nonumber\\
                     &&+ \frac{\Omega}{4\pi}
                         \int^{\infty}_{k_{fv}} dk \beta^2 k^7
                         \left( \frac{|v_{\mathbf{k}}|^2}{\varepsilon_{32}}
                              + \frac{|u_{\mathbf{k}}|^2}{\varepsilon_{42}}
                         \right)
                         \left( 1+ \cos^2 \theta_v \right) \nonumber\\
                     &&+ \frac{\Omega}{4\pi}
                         \int^{\infty}_{0} dk
                            \left[
                               \frac{\alpha^2 k^3 |v_{\mathbf{k}}|^2}{\varepsilon_{14}}
                               \left( 1+ \cos^2 \theta_c \right)
                              +\frac{\beta^2 k^7 |u_{\mathbf{k}}|^2}{\varepsilon_{24}}
                               \left( 1+ \cos^2 \theta_v \right)
                            \right. \nonumber\\
                       && \qquad\qquad\qquad
                            \left.
                              +\frac{\alpha^2 k^3 |u_{\mathbf{k}} v_{\mathbf{k}}|^2}{\varepsilon_{34}}
                               \left( 1 - \cos^2 \theta_c \right)
                              +\frac{\beta^2 k^7 |u_{\mathbf{k}} v_{\mathbf{k}}|^2}{\varepsilon_{34}}
                               \left( 1 - \cos^2 \theta_v \right)
                            \right] \nonumber\\
                      &=& \delta E^{(2)} + A \alpha^2 \cos^2 \theta_c + B \beta^2 \cos^2 \theta_v~.
\end{eqnarray}
Spin-rotational symmetry is now broken and depending on the sign of the constants $A$ and $B$,
The total spin can have either easy plane or easy axis.

\end{appendix}



\end{document}